\documentclass{aa}  
\usepackage{graphicx}
\graphicspath{ {figs/} }
\usepackage{color}
\usepackage{rotating}
\usepackage{hyperref} 
\usepackage{tablefootnote}
\usepackage[dvipsnames]{xcolor}
\usepackage{longtable}
\usepackage{pdflscape}
\usepackage{tikz}
 
\usepackage{pifont}
\usepackage[normalem]{ulem}
\usepackage{amsmath}

\newcommand{\struutup}{\rule{0ex}{3.2ex}}
\newcommand{\struutdown}{\rule[-2ex]{0ex}{2ex}}

\usepackage[varg]{txfonts}
\newcommand{\kms}{km\,s$^{-1}$~}
\newcommand{\kmsi}{km\,s$^{-1}$}
\newcommand{\Kepler}{\textit{Kepler~}}

\bibpunct{(}{)}{;}{a}{}{,}
\begin{document}
   \authorrunning{A. Samadi Ghadim, et~al.}
   \titlerunning{KIC~6951642: a \Kepler $\gamma$ Dor\,--\,$\delta$ Sct star in a possible single-lined binary system}
   \subtitle{KIC~6951642: confirmed \Kepler $\gamma$ Doradus\,--\,$\delta$ Scuti star with intermediate to fast rotation in a possible single-lined binary system 
   }
   \author{
          A. Samadi-Ghadim\inst{1 \href{https://orcid.org/0000-0002-1577-010X}{\includegraphics[scale=.02]{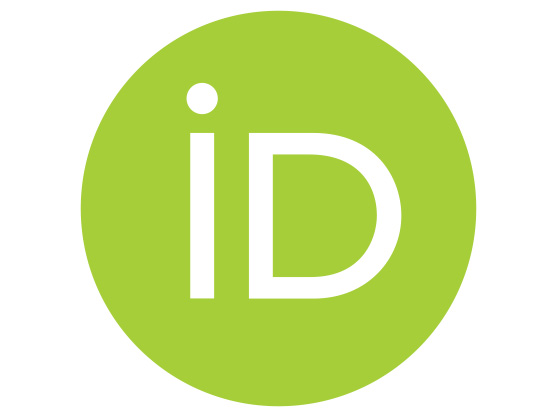}}}
          \and
          P. Lampens\inst{2 \href{https://orcid.org/0000-0002-7034-4912}{\includegraphics[scale=.02]{orcid.jpg}}}
         \and
          L. Gizon\inst{1,3,4 \href{https://orcid.org/0000-0001-7696-8665}{\includegraphics[scale=.02]{orcid.jpg}}}
         }
   \institute{Max-Planck-Institut für Sonnensystemforschung, 37077 Göttingen, Germany
   \email{samadi@mps.mpg.de}
   \and
   Koninklijke Sterrenwacht van Belgi\"e, Ringlaan 3, B-1180 Brussel, Belgium
   \and
   Institut für Astrophysik, Georg-August-Universität Göttingen, 37077 Göttingen, Germany
   \and
   Center for Space Science, NYUAD Institute, New York University Abu Dhabi, Abu Dhabi, United Arab Emirates}
   \abstract{KIC~6951642 has been reported as a candidate hybrid  pulsator of type-$\gamma$ Doradus \,--\, $\delta$ Scuti from observations of the first quarters of the \Kepler mission. The regular patterns seen in the Fourier spectra of the \Kepler and Transiting Exoplanet Survey Satellite TESS light curves and the sinusoidal modulation of its surface brightness suggest the additional presence of rotational modulation and stellar activity, respectively.}
   {We aim to investigate the pulsating nature of KIC~6951642 and to search for the signature of rotation and/or activity in the light curves.}
   {We performed an iterative frequency search of both Fourier spectra, and searched for regular patterns in them. We applied spectrum synthesis to determine the atmospheric stellar parameters. Since KIC~6951642 was reported to belong to a spectroscopic binary system, we fitted the time delays derived from the light curves with the radial velocities obtained from published as well as new spectra in an attempt to improve the quality of the first orbit.}
   {Follow-up spectroscopy showed that KIC~6951642 is a fast-rotating F0-type star in a possible single-lined binary with a period of $\sim$4.8~yr. In the low-frequency regime, we identified the frequencies of 0.721 d$^{-1}$~as well as of 0.0087 d$^{-1}$. We attribute the first frequency to stellar rotation, and the second one to stellar activity with a cycle of length of 3.2~yr. We also detected $g$ modes, with the strongest mode located at 2.238 d$^{-1}$, as well as three asymmetric multiplets (with a mean spacing of 0.675$\pm$0.044 d$^{-1}$). In the high-frequency regime, we detected frequencies of type-$\delta$ Scuti, with the strongest mode located at 13.96 d$^{-1}$, as well as seven asymmetric multiplets (with a mean spacing of 0.665$\pm$0.084 d$^{-1}$). We subsequently identified a few more frequencies that appear to be combinations of a $g$ or $p$ mode and one of the higher cited frequencies not due to pulsations.}
   {We propose that KIC~6951642 accomadates for a fast-rotating $\gamma$ Dor \,--\, $\delta$ Sct hybrid star with various rotationally split multiplets of $g$ and $p$ modes and that it also displays a cycle lasting years of (possible) stellar activity.}{} 
\maketitle
\section{Introduction}\label{sec:intro}
\begin{table*}
 \centering
	\caption{Summary of previous studies concerning hybrid $\gamma$ Dor\,--\,$\delta$ Scuti pulsations in binary stars. }
	\label{tab:prev_study}
	\begin{tabular}{lccccccccc}
	\hline\hline
   	ID &Type$^{(a)}$ & e$^{(b)}$& Period & $v\sin{i_{1,2}}$ & M$_{1,2}$ & $g$ modes& $p$ modes & Act.$^{(c)}$  \struutup\\
       &             &          &  d     & \kms             & M$_{\sun}$& d$^{-1}$ & d$^{-1}$  & Cycle (d)           \\
   \hline
  KIC~4544587$^{(1)}$   & EB   & 0.29 & 2.19  & 86.5        & 1.98             & $\gamma$ Dor \& tidal$^{(d)}$ & $\delta$ Sct      & No \struutup\\
                                 &      &      &       & 75.8        & 1.60             & 0.04-4.57                     & 38.2-48.05        &  \struutdown\\
  KIC~10080943$^{(2)}$  & EB   & 0.45 &15.34  & 19.0        & 2.0              & rot.-split$^{(e)}$            & rot.-split        & No\\             
                                 &      &      &       & 18.7        & 1.9              & 0.6-1.45                      & 12-20             &  \struutdown\\ 
  KIC~6048106$^{(3)}$   & EB   & 0.01 & 1.56  & -           & 1.55             & $\gamma$ Dor                  & tid.-split$^{(f)}$& Spot \\             
                                 &      &      &       & -           & 0.33             &  1.96-2.85                    & 7.49-15.2         & 290  \\  
                                 &      &      &       &             &                  &                               & 19-22.5           &      \struutdown\\ 
  KIC~4142768$^{(4)}$   & EB   & 1.0  & 13.99 & 8.67$^{(g)}$& 2.05             & $\gamma$ Dor \& tidal         & $\delta$ Sct      & No  \\
                                 & SB2  &      &       & 7.35$^{(g)}$& 2.05             & 0.1-3.0                       & 15-18             &      \struutdown\\ 
  KIC~8975515$^{(5)}$   & SB2  & 0.41 & 1603  & 162         &  0.83 (q$^{(h)}$) & rot.-split                    & rot.-split        & No  \\ 
                                 &      &      &       & 32          &                  & 1.56-6.18                     & 7.2-21.2          &     \struutdown\\
  TIC~11491822$^{(6),(i)}$& EB & 0.078& 9.94  &-            & 1.94             & $\gamma$ Dor \& tid.-split    & $\delta$ Sct      & No  \\ 
                                 &      &      &       &-            & 1.51             & 0.11-3.51                     &  5.09-18.9        &     \struutdown\\
  KIC~9850387$^{(7)}$   & EB   & 0.0  &  2.74 & 13.4        & 1.66             & $\gamma$ Dor \& tid.-split    & $\delta$ Sct      & No  \\
                                 &      &      &       &-            & 1.06             & 0.11-3.51                     & 10.7-16.6         &     \struutdown\\
   \hline \hline
   \end{tabular}
     \tablebib{(1) \citet{Hambleton2013} (2) \citet{Schmid2015} and \citet{Keen2015} (3)\citet{Samadi2018a} and \citet{Samadi2018b}, \citet{Lee2016} (4) \citet{Guo2019} (5) \citet{Samadi2020} (6) \citet{Southworth2021} (7) \citet{Sekaran2020}, \citet{Zhang2020} and \citet{Sekaran2021}}
     \tablefoot{\\
     \tablefoottext{a}{Type i.e. binary classification: EB: Eclipsing Binary, SB2: double-lined Spectroscopic Binary.}
     \tablefoottext{b}{e: eccentricity}
     \tablefoottext{c}{Act.: any signature of stellar {\bf Act}ivity}
     \tablefoottext{d}{tidal: {\bf tidal}ly excited modes}
     \tablefoottext{e}{rot.-split: {\bf rot}ationally{\bf split} modes}
     \tablefoottext{f}{tid.-split: {\bf tida}lly{\bf split} modes }
     \tablefoottext{g}{$v\sin{i}$ from model}
     \tablefoottext{h}{q = M$_{2}$/M$_{1}$: the mass ratio} 
     \tablefoottext{i}{RR Lyn}
     }
\end{table*}
Main-sequence A- and F-type stars are intermediate-mass stars with luminosities in the range 43 - 2 L$_{\sun}$ and effective temperatures (T$_{\rm eff}$) between 9800 and 6000~K \citep{Cox2000}. As T$_{\rm eff}$ drops beyond $\sim$7000~K and the sudden onset of convection starts \citep{Christensen2000, DAntona2002}, such stars are in a critical phase of transition as the regime of energy transfer in their stellar envelopes changes from (mainly) radiative to convective. This structural modification also leads to different mechanisms that can generate stellar pulsations. In this part of the H-R diagram, the pulsating stars comprise the $\gamma$ Doradus (Dor), the (low- and high-amplitude) $\delta$ Scuti (Sct) stars, as well as the rapidly oscillating (magnetic) Ap stars. A detailed description of the properties of these pulsators can be found in the works by \citet{Aerts2010}, \citet{Balona2014} and \citet{Antoci2019}, while a review of the recent discoveries concerning them based on the planet-searching space missions is provided by \citet{Antoci2019}. Here, we provide a brief overview of the general properties of only two groups.\\ 
The mid-F- to late-A-type $\gamma$ Dor stars (with masses of 1.4-1.9 M$_{\sun}$) pulsate in high-order, low-degree $g$ modes with periods of 7~hr to 3~d \citep{Kaye1999}. The $g$ modes are excited by a flux modulation mechanism at the base of the convective zone \citep{Guzik2000,Dupret2004,Dupret2005,Grigahcene2010}.\\ 
The early-F-type to early-A-type $\delta$ Sct stars (with masses of 1.5-2.5 M$_{\sun}$) pulsate in low-order radial and non-radial pressure modes ($p$ modes) with typical frequencies in the range (4 or 5) - 65 d$^{-1}$. These modes are mainly driven by the $\kappa$ mechanism operating in the partial ionisation zone of He~II \citep{Gautschy1995,Breger2000}, as well as by turbulent pressure which is responsible for the excitation of $p$ modes of moderate radial order \citep{Antoci2014,Xiong2016}. Such modes allow one to probe the upper stellar layers. In unevolved $\delta$ Sct stars, most of the excited modes are $p$ modes. However, as the star evolves and the frequency spectrum becomes denser, a large range of $p$ modes, mixed modes \citep{Aizenman1977, chen2018} and $g$ modes is predicted to be excited \citep{Handler2013}, that is to say $g$ modes such as those found in $\gamma$ Dor stars \citep[e.g.][]{Balona2011}. \\
The space missions, for example \Kepler and Transiting Exoplanet Survey Satellite TESS \citep{Ricker2014,Ricker2015}, revealed a new class of pulsators for which both mode-driving mechanisms operate simultaneously, that is to say the hybrid $\delta$ Sct \,--\,$\gamma$ Dor stars and hybrid $\gamma$ Dor\,--\,$\delta$ Sct stars. Such hybrid pulsators cover a region of the H-R diagram far more extended than the theoretical overlapping zone of the instability strips associated with each class \citep{Balona2015,Bowman2018}. The simultaneous excitation of $g$  and $p$ modes poses a big problem from a theoretical point-of-view. There is an on-going discussion about what causes $g$ modes in the hottest $\delta$ Sct stars \citep{Balona2011,Balona2015,Xiong2016,Lampens2018}. From asteroseismic studies of hybrid pulsators, we can infer the stellar rotation and chemical profiles from the near-to-core to the surface layers \citep[e.g.][]{VanReeth2018,Li2020}. \\
\citet{Duchene2013} reported an observed frequency of spectroscopic binaries of the order of 30-45\% among intermediate-mass field stars. A recent study by \citet{Moe2017} shows that the binary star fraction among A/late B stars (M$_{1}$ = 2-5 M$_{\sun}$) is 37$\pm$6\%. Thus, we may expect a significant fraction of binaries among intermediate-mass (field) pulsators. A binary fraction of 14 $\pm$ 2\% was derived by \citet{Murphy2018} from a sample of 2200 \Kepler main-sequence A/F-type pulsators. Based on multi-epoch spectroscopy, \citet{Lampens2018} reported a multiplicity fraction of at least 27\% in a sample of 49 \Kepler hybrid pulsators of similar spectral type. As material of comparison for our study, we provide a summary of recent studies of A/F-type hybrid pulsators in binary systems in Table~\ref{tab:prev_study}. In this table, we report some orbital and stellar parameters, the range of observed frequencies, and the origin of the frequency splittings detected among the $g$  and $p$ modes. Some of these pulsators in eclipsing binaries were also discussed by \citet{Lampens2021b}. \\
In this study, we are concerned about KIC~6951642, a \Kepler candidate for hybrid pulsations in a possible long-period binary system. We aim to decode the nature of the dominant frequencies in both the low- and high-frequency regions of its Fourier spectrum derived from \Kepler's full range observations. \\
The structure of this paper is the following: Section~\ref{sec:intro} presents the general context. Section~\ref{sec:prev&spec} describes the current status based on the literature as well as follow-up spectroscopy. We explain how we derived the atmospheric stellar properties and revised the (preliminary) orbital solution applying a combined least-squares fitting of radial velocities (RVs) and time delays (TDs). We compare the light curves available from the different surveys and describe their preparation for the study of KIC~6951642 in Sect.~\ref{sec:LC}. In Sect.~\ref{sec:activity}, we report our attempt concerning the search for evidence of stellar activity. We next provide the results based on the photometric (Sect.~\ref{sec:activity-photom}) and spectropolarimetric (Sect.~\ref{sec:activity-specpolar}) observations of KIC~6951642. Section~\ref{sec:fourier-comb} details and compares the most significant features of the Fourier spectra obtained from the \Kepler and TESS data as well as concerning the detection of combination and harmonic frequencies in the complete frequency range. We report our results concerning the analyses of the low-frequency region and detection of the rotationally split $g$ modes (in Sect.~\ref{sec:low-freqs}). In Sect.~\ref{sec:high-freqs}, we carefully search for $\delta$ Scuti pulsations and any signature of rotationally split $p$ modes in the high-frequency region of the Fourier spectra. Finally, we discuss and summarise the main results of this study in Sect.~\ref{sec:conclusion}.
\section{Follow-up spectroscopy}\label{sec:prev&spec}
\begin{figure}
\centering
	\includegraphics[width=\columnwidth]{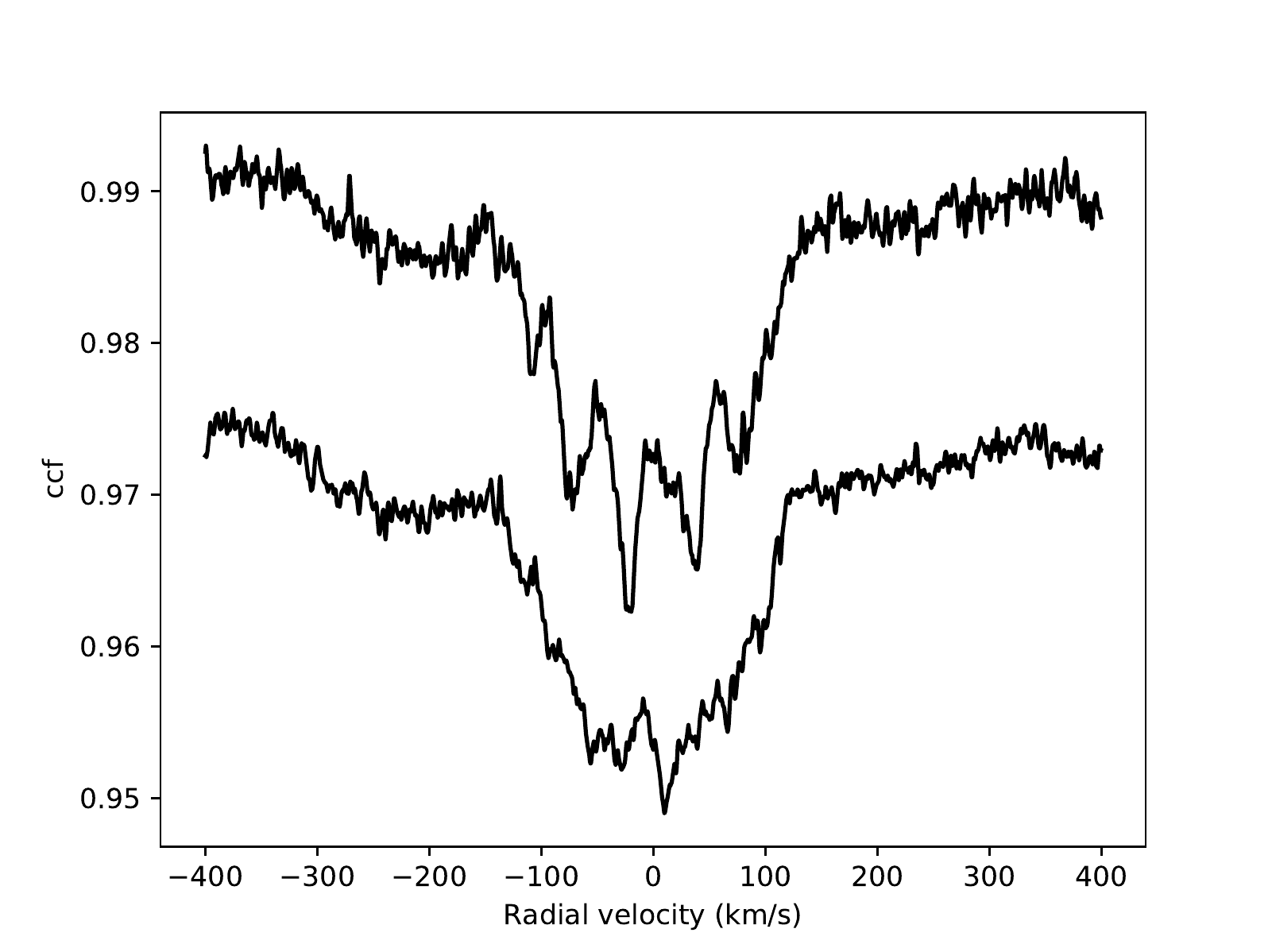}
	\caption{CCFs from \textsc{HERMES} spectra collected during two different nights.}
	\label{fig:ccfs}
\end{figure}

KIC~6951642 was classified as a new hybrid star in the characterisation study of the pulsational behaviour of 750 A-F-type stars observed by \Kepler \citep{Uytterhoeven2011}. In this study based on data of Q0-Q1 (cf. their Table~3), this star presents a comparable number of low and high frequencies with the most dominant one located at 0.721 d$^{-1}$. It was also reported as a probable hybrid star showing both $\gamma$ Dor and $\delta$ Sct oscillations by \citet{Fox-Machado2017}. As a follow-up of the study by \citet{Uytterhoeven2011}, \citet{Lampens2018} included this object among a list of 50 hybrid candidate stars to be surveyed with the high-resolution, high-efficiency spectrograph \textsc{HERMES} \citep{Raskin2011}. From their multi-epoch spectroscopic survey, KIC~6951642 was classified as 'P+VAR'. \citet{Lampens2018} assigned it to class 'P' because the cross-correlation functions CCF display strong line-profile variations, either due to non-radial stellar pulsations (P) or to stellar rotation in the presence of surface brightness variability or spots (ROT) (Fig.~\ref{fig:ccfs}). The plot of the radial velocities (RVs) against time (Fig.~\ref{fig:rvs}) indicates low-amplitude variability over a long period which is the reason for its supplementary classification 'VAR'. Since such low-amplitude, long-term variations could also be caused by binarity, a follow-up study was necessary (cf. Sect.~\ref{sec:orb-new}). 
\begin{figure*}
\centering
	\includegraphics[width=\columnwidth]{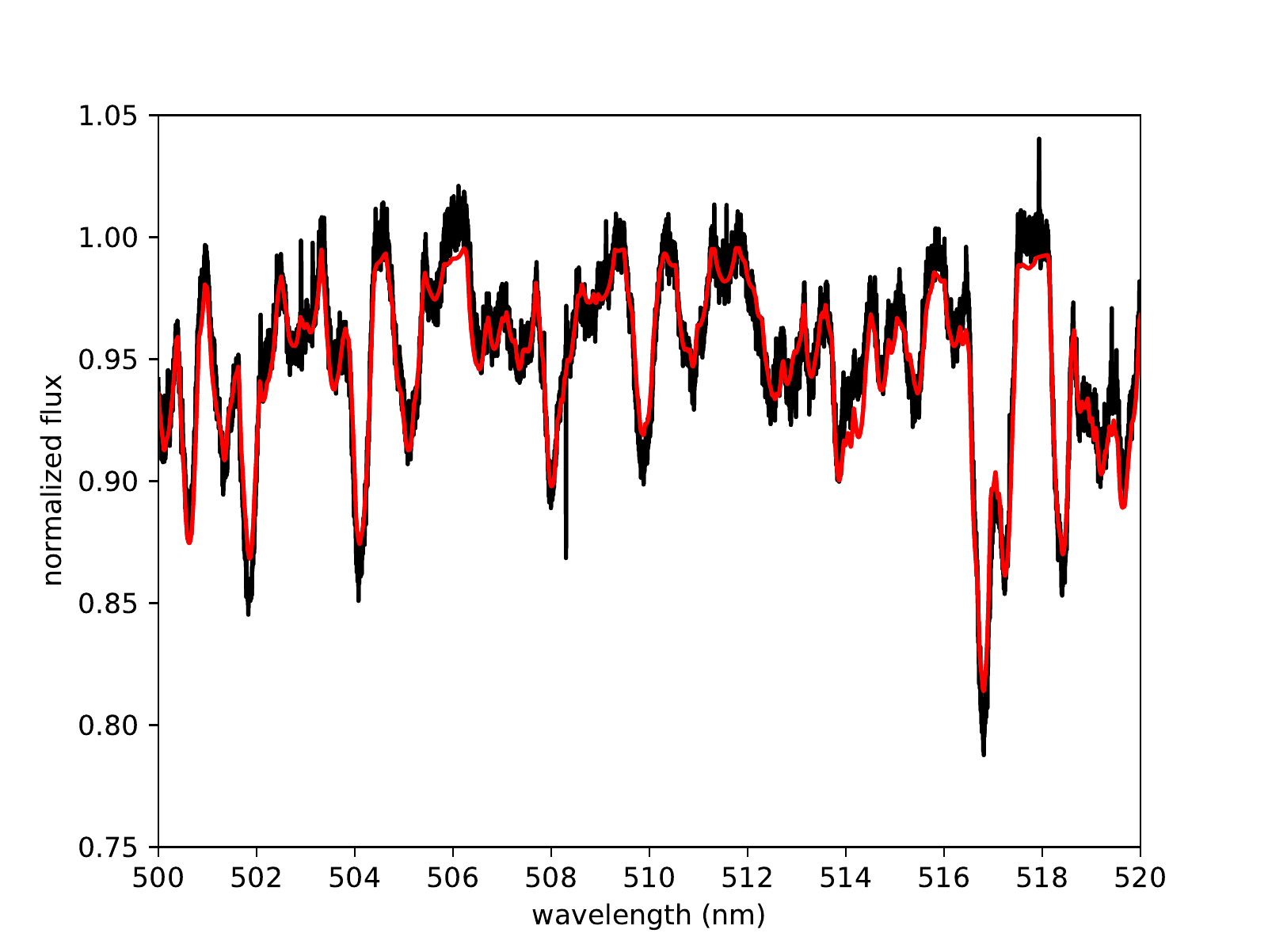}
	\includegraphics[width=\columnwidth]{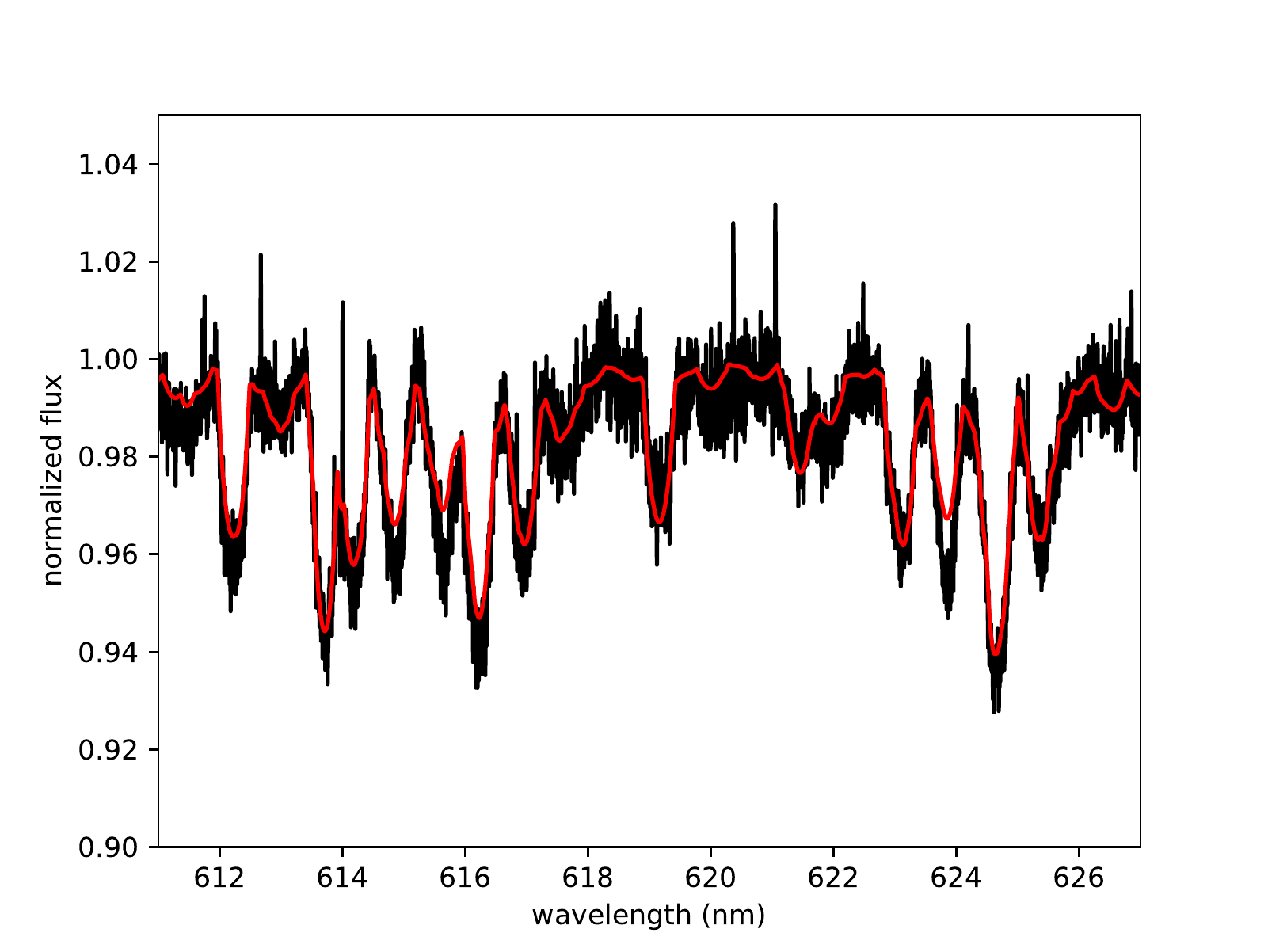}
	\caption{Part of the \textsc{HERMES} sum spectrum (black) and its matching synthetic spectrum (red) {\it Left:} in the range 500-520 nm. {\it Right:} in the range 600-630 nm.}
	\label{fig:s_spectrum}
\end{figure*}
\begin{figure}
\centering
	\includegraphics[width=\columnwidth]{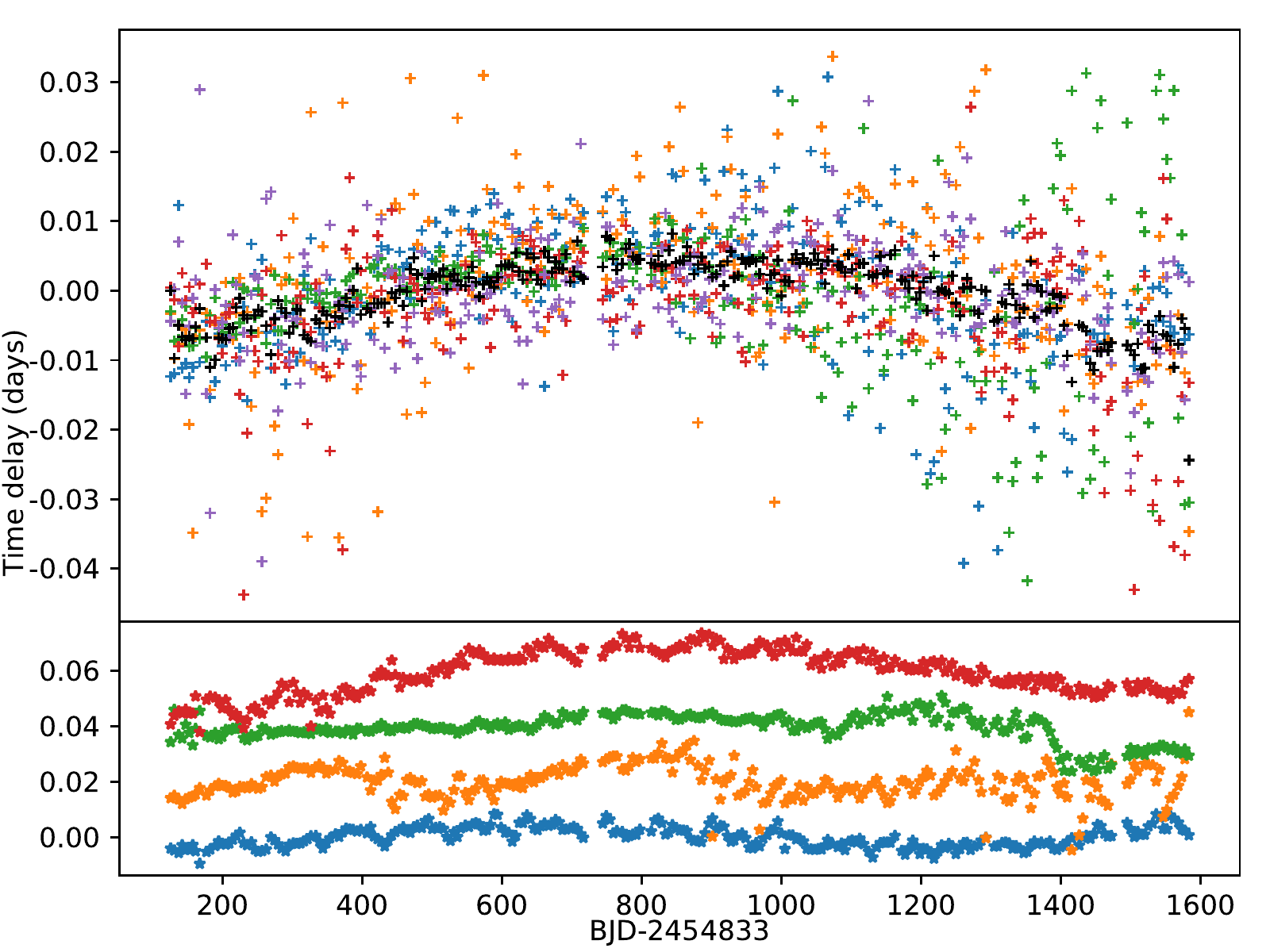}
	\caption{TDs of high-frequencies. Top panel: eight frequencies displaying a similar long-term change: $f_{39}$ = 13.0064 d$^{-1}$ ($f_{\rm parent}$), $f_{85}$ = 10.3107 d$^{-1}$ ($f_{\rm comb}$), $f_{107}$ = 15.4706 d$^{-1}$ ($f_{\rm parent}$), $f_{123}$ = 14.8528 d$^{-1}$ ($f_{\rm parent}$), $f_{126}$ =12.7318 d$^{-1}$ ($f_{\rm parent}$), $f_{146}$ = 12.5150 d$^{-1}$ ($f_{\rm parent}$), and $f_{178}$ = 13.7274 d$^{-1}$, $f_{43}$ = 14.3778 d$^{-1}$. Their mean values are plotted in black. Bottom panel: three frequencies whose behaviour is incompatible with the general trend: $f_{31}$ = 13.9651 d$^{-1}$ ($f_{\rm p_{\rm max}}$, in blue), $f_{59}$ = 14.9697 d$^{-1}$ ($f_{\rm parent}$, in orange) and $f_{44}$ = 15.9634 d$^{-1}$ ($f_{\rm parent}$, in green), and $f_{178}$ = 13.7274 d$^{-1}$, $f_{43}$ = 14.3778 d$^{-1}$. We re-plotted $f_{43}$ = 14.3778 d$^{-1}$ ($f_{\rm parent}$, in red) in the lower panel for means of comparison with the frequencies that have incompatible trends in top panel. See Table~\ref{tab:terminology} for the used terminology. }
	\label{fig:tds46fs}
\end{figure}
\begin{figure}
\centering
	\includegraphics[width=\columnwidth]{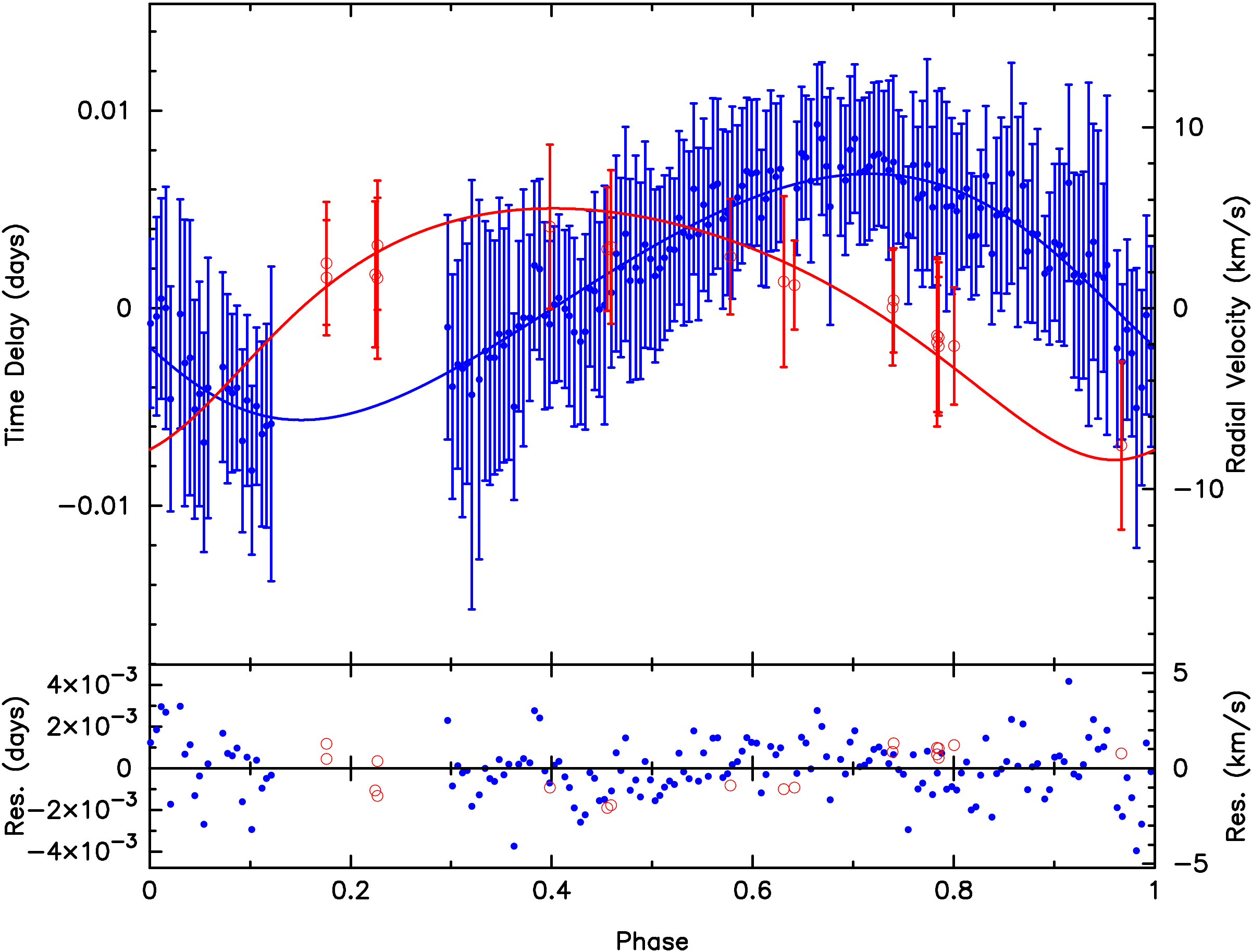}
	\caption{Orbital solution for KIC~6951642 based on the \textsc{HERMES} RVs (red symbols and red solid line) and the weighted mean TD values (blue symbols and blue solid line) of eight high frequencies derived from the \Kepler data. }
	\label{fig:combined_solution}
\end{figure}
\begin{table}
 \centering
	\caption{Best-fitting set of atmospheric properties from the normalised \textsc{HERMES} sum spectrum of KIC~6951642.}
	\label{tab:atm_par}
	\begin{tabular}{lcccc}
	\hline\hline
   	Parameter & Value \\
   \hline
   $v \sin{i}$ (\kms)      & 123$\pm$3 \struutup\\
   T$_{\rm eff}$ (K)       & 7336$\pm$186   \\
   $\log{g}$ (cgs)         & 3.94$\pm$0.26  \\
   $[\rm{M}/\rm{H}]$ (dex) & -0.17$\pm$0.08  \struutdown\\
   \hline
   \end{tabular}
\end{table}
\begin{table}
\center
\begin{minipage}{8.9cm}
\centering \caption[]{\label{tab:vop6} Values and standard deviations of the constrained parameters of the updated orbital solution proposed for KIC~6951642. }
\begin{tabular}{lll}
\hline
\hline
\multicolumn{3}{c}{Combined orbital solution A-B}\\
\hline
Parameter &\multicolumn{1}{l}{Value} & \multicolumn{1}{c}{$\sigma^{(a)}$}\\
\hline
P$_{\rm orb}$ (d)               &  1796   & $^{+55}$/$_{-69}$ \struutup\\
T$_{\rm{0}}$ (HJD$^{(b)}$) & 2457904 & $^{+170}$/$_{-153}$ \\
e                   &  0.25   & $^{+0.14}$/$_{-0.18}$ \\
$\omega$ (\degr)      &   187   & $^{+36}$/$_{-31}$  \\
V$_0$ (\kmsi)         & -3.4    &   1.2 \\
K$_{\rm{A}}$ (\kmsi)  &  7.2    &   0.9 \\
a$_{\rm{A}} \mathrm{sin}\,i$ (AU)    & 1.14  & 0.12   \\
a$_{\rm{TD}}\,\mathrm{sin}\,i$ /c (d) & 0.0066 & 0.0007 \\
f(${\rm{M}}_{\rm{B}}$) (M$_\odot$)     & 0.06  & 0.02  \struutdown\\
\hline
rms$_{\rm{A}}$ (\kmsi) &1.33  &  \struutup\\
rms$_{\rm{TD}}$ (d)    &0.002 &  \struutdown\\
\hline
\end{tabular}
     \tablefoot{\\
     \tablefoottext{a}{$\sigma$: Standard deviation}
     \tablefoottext{b}{HJD: Heliocentric Julian Date}
     }
\end{minipage}
\end{table}
\subsection{Atmospheric properties}\label{sec:orb_prev}
We acquired 21 \textsc{HERMES} spectra of KIC~6951642 over a period of 10 years (see Table~\ref{tab:journal} for detailed information about the spectra). Based on these data, we re-derived the atmospheric stellar properties from the normalised \textsc{HERMES} sum spectrum. This spectrum was evaluated against a grid of synthetic spectra to obtain the best fitting set of parameters using the tool 'iSpec' from \citet{Blanco-Cuaresma2014} and \citet{ Blanco-Cuaresma2019}. In this comparison, we used the metal lines only. We list the corresponding set of parameters in Table~\ref{tab:atm_par}. Figure~\ref{fig:s_spectrum} illustrates the excellent match between the observed and the synthetic spectrum in two very different wavelength ranges. The conclusion is that the spectrum of \object{KIC~6951642} corresponds to that of a normal F0-type star. There is no trace of any additional contribution in the spectrum.
\subsection{KIC~6951642 as a possible SB1: combined orbital solution}\label{sec:orb-new}
One possible mechanism to explain the slow, low-amplitude change of the RVs could be orbital motion in a binary system. \citet{Murphy2018} reported time delay variability for nine frequencies with a possible period longer than the duration of the \Kepler data. From a simultaneous modelling of their time delays (TDs) and \citet{Lampens2018}'s RVs, they classified the target as a long-period spectroscopic binary with P$_{\rm orb}$ = 1867~d, the semi-amplitude of the light-travel time, a$_{\rm TD} \sin {i/\rm c}$ = 458$\pm$9 s and the corresponding velocity semi-amplitude, K$_{\rm A}$ = 6.1$\pm$0.2 \kmsi. \\ 
We tried to improve the orbital solution of KIC~6951642 by performing a similar modelling based on the larger set of the RVs (Col.~4 of Table~\ref{tab:journal}) and the TDs since the origin of the TD variability could be identical to that of the RV variability \cite[see for example][]{Lampens2021b}. \\
First, we recomputed the TDs for many frequencies of the \Kepler Fourier spectrum (Fig.~\ref{fig:pergrams}), in particular for the dominant frequencies of the high-frequency region (where the $p$ modes are found), but it was not possible to find a consistent behaviour for all the considered frequencies. In the next step, we selected the frequencies whose observed TDs show a long-term behaviour compatible with that of the RVs. The top panel of Fig.~\ref{fig:tds46fs} displays the TDs of eight frequencies that show a similar general trend, whereas the bottom panel of Fig.~\ref{fig:tds46fs} displays TDs of three other frequencies clearly showing a different trend on the long term and/or chaotic behaviour (in some parts). We re-plotted $f_{43}$ (= 14.3778 d$^{-1}$) in red in the bottom panel for means of comparison with the incompatible trends. By using these data together with the RVs in a simultaneous least-squares analysis, we obtained a revised solution, with the orbital parameters of Table~\ref{tab:vop6}. The revised orbital solution has P$_{\rm orb}$ $\sim$ 1796~d, a$_{\rm TD}\sin{i}/c$ = 0.0066~d (570~s) and K$_{\rm A}$ = 7.2$\pm$0.9~\kmsi. The \textit{rms} of the RV residuals is 1.33~\kmsi, that is to say larger than expected for a single-lined system (SB1) of spectral type F0, while that of the TDs is 0.002~d (Fig.~\ref{fig:combined_solution}). Due to the long period and the uncertainties, the coverage of the RV curve is far from complete. Follow-up RVs are definitely needed in the next couple of years to confirm the binary nature of KIC~6951642. The reason why some dominant pulsation frequencies fail to show a consistent long-term trend in their TDs also needs to be clarified. Some modes may present intrinsic phase changes on top of the light travel time effect (LITE). It is possible that such intrinsic variability does not allow us to find a common trend for all the $p$ modes.
\section{Available photometric data}\label{sec:LC}
\begin{table}
 \centering
	\caption{Information from literature, surveys and large data bases.}
	\label{tab:photom_data}
	\begin{tabular}{lccc}
	\hline\hline
	{\textbf Survey}      & {\textbf Target ID} \\
	\hline
	\emph{\Kepler}        & KIC~6951642  \struutup\\
	TESS                  & TIC~63371872\\
	Gaia                  & 2125743657421198592\struutdown \\
	\hline
	KIC$^{(a)}$\\
	\hline
	RA                     &  19$^{h}$31$^{m}$05.93$^{s}$ \struutup\\
	Dec                    & +42$\degr$29$\arcmin $53.20$\arcsec$\\
    Kp (mag)               & 9.70  \\
	T$_\mathrm{eff}$ (K)   & 7178  \\
	$\log{g}$ (cgs)        & 3.365  \\
    R (R$_{\odot}$)        & 4.413\struutdown\\               
    \hline
    Gaia DR2$^{(1)}$\\
    \hline
     $\varpi$ (mas)    & 1.7272$\pm$0.0257 \struutup\\
     distance$^{(2)}$(pc)  & 569.6085 \\
     m$_{\rm G}$ (mag) & 9.60 \struutdown\\
     \hline
    Gaia DR3$^{(3)}$\\
    \hline
    $\varpi$ (mas)     & 1.7077$\pm$0.0153 \struutup\\
     m$_{\rm G}$ (mag) & 9.61 \struutdown\\
   \hline
    \end{tabular}
    \tablebib{(1)~\citet{Gaia2016b}, \citet{Gaia2018a} and \citet{Gaia2018b}
    (2)~\citet{Bailer-Jones2018A}(3)~\citet{Gaia2021}, \citet{GaiaEDR32021} and \citet{Gaia-astrometric2022}}
    \tablefoot{\\
    \tablefoottext{a}{from the \Kepler Input Catalogue (based on 5-band photometry, plus the J, H, \& K bands from the 2MASS survey, assuming that the object is a single star.)}
    }
\end{table}
\begin{table}
 \centering
	\caption{Available photometric data for KIC~6951642. }
	\label{tab:LC_info}
	\begin{tabular}{lcccccc}
	\hline
	\hline
	Parameter/Survey             & \Kepler LC & \Kepler SC & TESS $^{(b)}$\\
	\hline\struutup
    T (d)                     & 1470.462  & 9.726   & 25.979 \\
    $\Delta$t (min)           & 29.424    & 0.981   & 29.999 \\
    T$_{\rm ref}$ (BJD)       & 2454833   & 2454833 & 2457000\\
    $f_{\rm res}$ (d$^{-1}$)  & 0.00068   & 0.10282 & 0.03849\\
    $f_{\rm Nyq}$ (d$^{-1}$)  & 24.469    & 734.074 & 24.001 \\
    interval                  & Q0-Q17    & Q0      & S15   \\
    year                      & 2009-2013 & 2009    & 2019  \struutdown\\
	\hline
	\end{tabular}
	\tablefoot{T: Time span of the observations. $\Delta$t: mean time sampling of the observations. $f_{\rm res}$: Resolution frequency. $f_{\rm Nyq}$: Nyquist frequency.\\
	\tablefoottext{b}{pipeline: TESS Light Curves From Full Frame Images (TESS-SPOC)}}
\end{table}
\begin{figure*}
\centering
	\includegraphics[width=16cm, height=9.5cm,keepaspectratio]{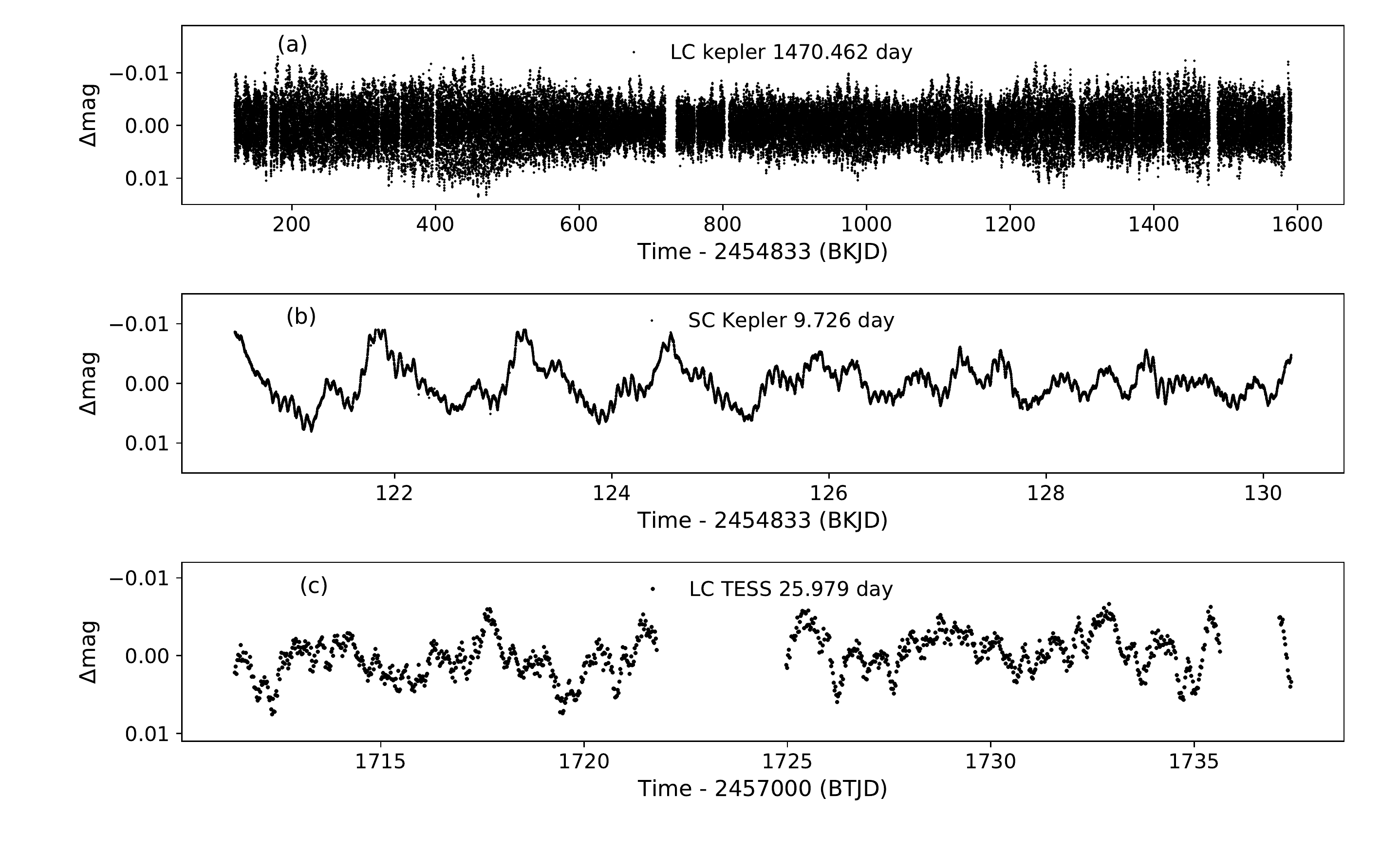}
	\caption{Observed Light curves for KIC~6951642 from (a) \Kepler~Long Cadence (LC) (b) \Kepler~Short Cadence (SC) (c) TESS 1800-s sampling, sector 15 (Table~\ref{tab:LC_info}).}
	\label{fig:lightcurves}
\end{figure*}
 KIC~6951642 is a bright \Kepler object K$_{\rm p}$ = 9.70 mag (Table~\ref{tab:photom_data}). Figure~\ref{fig:lightcurves} shows all available light curves for KIC~6951642. We reported the information of each light curve in Table~\ref{tab:LC_info}. The top panel in Fig.~\ref{fig:lightcurves} illustrates the full four-year \Kepler (1470.462 days) long-cadence (LC) observations with a sampling of 29.42 min. We normalised and stitched the light curves from different quarters with the python package 'LightKurve 2.0' \citep{Lightkurve2018}. We removed the outliers (with $\sigma > 3$) and converted the data from flux to relative magnitude. We used the outcome of this process, the 'detrended' LC light curve, in our study. For the observed light curves in Table~\ref{tab:LC_info}, we proceeded with the same detrending method. Figure~\ref{fig:lightcurves} shows that the relative magnitude varies with a semi-amplitude of 8 mmag on average in all panels. The TESS light curve (panel (c) of Fig.~\ref{fig:lightcurves}) with an 1800-s sampling rate shows the smallest semi-amplitude of all data sets (7 mmag). We can see a strong repetitive structure during the first days of the short light curve. The contamination factors reported by Kepler and TESS Input catalogues for KIC~6951642 are very small, that is to say 0.014 for KIC and 0.05 for TIC. We also compared the light curves derived directly from the target pixel files (TPF) for the \Kepler and TESS data (Table~\ref{tab:LC_info}) by applying a customised aperture mask smaller than the default aperture mask of respective pipelines (e.g. Figure~\ref{fig:lightcurves}). The light curves derived with the customised and the default masks are the same, that is to say both light curves show the same amplitudes of their variations. Hence, we conclude that the origin of the brightness variations is stellar rather than instrumental.
\section{Search for evidence of stellar activity}\label{sec:activity}
The relative surface brightness of the LC light curve (panel (a) of Fig.\ref{fig:lightcurves}) shows a noticeable increase at two intervals, that is to say the first $\sim$500 days (Q0-Q7) and the last $\sim$200 days of the observations (Q15-Q17). Indeed, KIC~6951642 shows very regularly spaced peaks in the low-frequency part of the Fourier spectrum of its \Kepler light curve (Sect.~\ref{sec:fourier-comb}), which could be indicative of the presence of a magnetic field. Hence, KIC~6951642 was a target of a spectropolarimetric survey to search for magnetic $\delta$ Scuti stars among potential candidates selected from the \Kepler data (Thomson-Paressant et al. \textit{in prep.}). In addition to the spectropolarimetric observations, we investigated the light curve in search of any signature of stellar activity, which we explain in this section.  
\subsection{Photometric observations}\label{sec:activity-photom}
The stellar surface brightness may experience long-term fluctuations due to a change in the location and the size of the stellar spots \citep{Garcia2010}. This allows us to detect the activity cycle by measuring the photospheric magnetic activity proxy S$_{\rm ph}$ from the light curve \citep{Mathur2014}. We derived S$_{\rm ph}$ using the full 4-year \Kepler LC light curve and plotted it versus the year of the observations in Figure~\ref{fig:magnetic_cycle}. We adopted the frequency of 0.721 d$^{-1}$ \citep{Uytterhoeven2011} assuming that it corresponds to the rotation frequency. Consequently, a sinusoidal behaviour of S$_{\rm ph}$ was observed allowing to distinguish phases of (relative) activity and inactivity. The frequency of this signal is $f_{\rm ac}$ = 0.00085$\pm$0.0002 d$^{-1}$, equivalent to $\sim$3.205 yr. We illustrate the best-fitting sinusoid to S$_{\rm ph}$ in orange in panel (b) of Fig.~\ref{fig:magnetic_cycle}. In panel (c) of Fig.~\ref{fig:magnetic_cycle}, we plotted the residuals emphasising their standard deviation, $\sigma$ = 500 $\mu$mag. The frequency corresponding to S$_{\rm ph}$ lies very close to both the resolution frequency of the \Kepler LC observations ($f_{\rm res}$ = 0.00068 d$^{-1}$) and the (unresolved) orbital frequency ($f_{\rm orb}$ = 0.00056$\pm$0.00002 d$^{-1}$).\\
\subsection{Spectropolarimetric observations}\label{sec:activity-specpolar}
Thomson-Paressant et al. (\textit{private commun.}) observed KIC~6951642 with ESPaDOnS at the Canada-France-Hawaii telescope (CFHT) of the Mauna Kea Observatory in Hawaii on three different nights (April 19 and 22, and May 14, 2016). There is no sign of a magnetic field on the analysed Stokes V data with the Least-Square Deconvolution (LSD). There is no detection of any Zeeman effect in any of the three observations. The three longitudinal field values are compatible with 0 within 2$\sigma$. From the low noise level of the ESPaDOnS data, Thomson-Paressant et al. (\textit{in prep.}) deduced that any dipolar magnetic field B $>\sim$300 G at the pole is detectable in the ESPaDOnS data with a probability of 90\%.
\begin{figure*}
\centering
	\includegraphics[width=16cm, height=9.5cm,keepaspectratio]{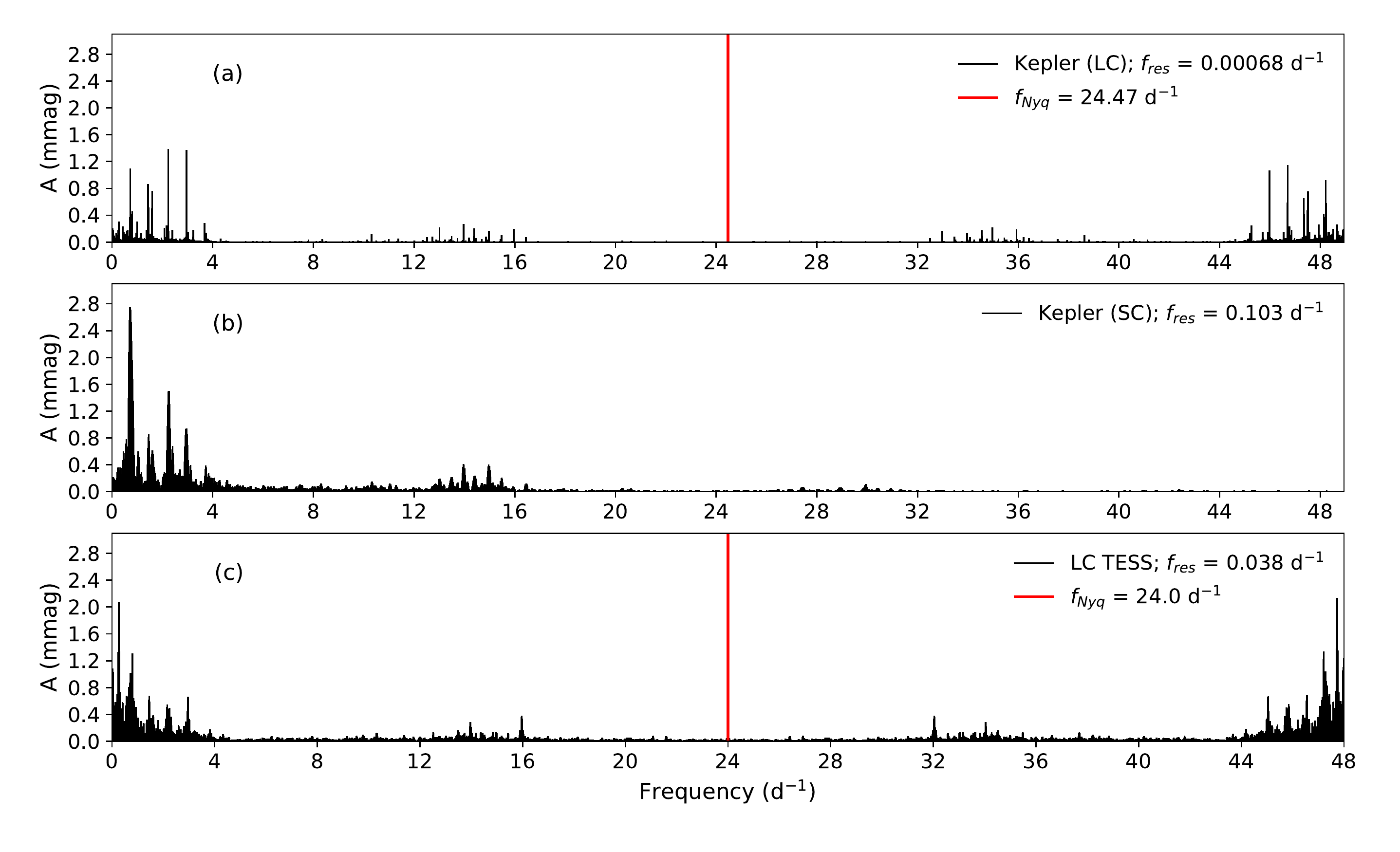}
	\caption{Fourier Spectra for KIC~6951642 from (a) \Kepler Long Cadence (LC) (b) \Kepler Short Cadence (SC) (c) TESS 1800-sec, sector 15. The spectra in (a) and (c) are calculated up to their double Nyquist frequencies.}
	\label{fig:pergrams}
\end{figure*}
\begin{table*}
 \centering
	\caption{Terminology used to describe the combination frequencies found in the Fourier spectrum of KIC~6951642 (Sect.~\ref{sec:fourier-comb}). }
	\label{tab:terminology}
	\begin{tabular}{llll}
	\hline\hline
	Frequency                  & Description & value (d$^{-1}$) & Colour \\
	\hline
	$f_{\rm ac}$               & the frequency of long-term stellar activity       & 0.00085$\pm$0.0002  &  -- \struutup\\
	$f_{\rm ach1}$             & $f_{117}\sim$ 9$^{th}$ harmonic of  $f_{\rm ac}$  & 0.0087              & \textcolor{PineGreen}{green}\\
	$f_{\rm ach2}$             & $f_{169}\sim$ 9$^{th}$ harmonic of  $f_{\rm ach1}$& 0.0856              & \textcolor{PineGreen}{green}\\
	$f_{\rm rot}$              & $f_{3}$, the candidate rotational frequency       & 0.721               & \textcolor{Red}{red}\\
	$f_{\rm parent}$           & the parent $g$  and $p$ modes                     &                     & \textcolor{magenta}{magenta}\\
	$f_{\rm p_{\rm max}}$      &  $f_{31}$, the highest amplitude $p$ mode         & 13.965              & \textcolor{violet}{violet}\struutdown\\
   \hline \hline
   \end{tabular}
   \tablefoot{The uncertainties in frequency are of the order of $\epsilon_{f}=(0.04-3) \times 10^{-4}$ d$^{-1}$. Col. 'Colour' refers to the colors in Tables~\ref{tab:frequencies_greigon} and \ref{tab:frequencies_preigon}.}
   \end{table*}
\section{Fourier analysis and combination frequencies}\label{sec:fourier-comb}
Our approach for Fourier analysis is the Lomb-Scargle periodogram \citep{Lomb1976, Scargle1982}. To determine the significant frequencies, we applied prewhitening, That is to say a sinusoid with the frequency of that highest peak in the periodogram is fitted to the light curve and subsequently subtracted from the light curve, iteratively. We calculated the Fourier spectrum of all three data sets (Table~\ref{tab:LC_info} \& Fig.~\ref{fig:lightcurves}) up to twice their Nyquist frequency, $f_{\rm Nyq}$ in Table~\ref{tab:LC_info}. The signal-to-noise ratio (S/N) of the frequencies is the mean of the periodogram over a window size of 1 d$^{-1}$. We adopted the much used criterion S/N $\geq$ 4 \citep{Breger1993} to consider a frequency as significant in this study. We consider that two frequencies are well resolved if their difference agrees with $f_{i} - f_{j} \geq 1.5/T$, with T the time span of the observation. $f_{\rm res}$ (or 1/T) for the \Kepler LC observations equals 0.00068 d$^{-1}$ (see Table~\ref{tab:LC_info} for the other observations). From each pair of frequencies that did not fulfil this criterion, we ignored the frequency of smaller amplitude. We obtained a list of 593 well-resolved frequencies from this procedure, called 'significant frequencies' hereafter. \\
Figure~\ref{fig:pergrams} shows the original Fourier spectra of the \Kepler LC (panel a), \Kepler SC (panel b) and TESS light curves (panel c). For the \Kepler LC and TESS spectra, we indicate the limiting Nyquist frequency with a vertical red line. We plotted a closer view of the Fourier spectrum of the \Kepler SC light curve in the range of the Nyquist frequency (panel (b) in Fig.~\ref{fig:pergrams}) to compare with the \Kepler LC spectrum. The larger frequency range in SC spectrum, that is to say 48-734 d$^{-1}$, doesn't include any significant frequency, so we cut that frequency region in panel (b) Fig.~\ref{fig:pergrams}. It is clear from the three spectra that there are two frequency ranges below the Nyquist regime where the frequencies with the largest amplitudes appear, that is to say at 0.2-4.0 d$^{-1}$ and 10-17 d$^{-1}$ intervals. In the super-Nyquist regime, the (still well resolved) mirrored frequencies appear with amplitudes either smaller than (\textit{Kepler}) or equal to (TESS) those of the already detected frequencies. To study the pulsations, we used the \Kepler LC Fourier spectrum. \\
A close look to the low- and high-frequency regions of the three Fourier spectra (Fig.~\ref{fig:pergrams_zoom}) confirms the presence of the most dominant frequencies in all of them. However, the amplitudes of the corresponding frequencies vary quite significantly from spectrum to spectrum. \\
The forest of frequencies found in hybrid ($\delta$ Sct\,--\,$\gamma$ Dor or $\gamma$ Dor\,--\,$\delta$ Sct) stars can include frequencies that are actually a linear combination of several independent frequencies with the highest S/N and largest amplitudes, named 'parent frequencies' hereafter. We marked the adopted parent frequencies as '$f_{\rm parent}$' (in magenta) in Tables~\ref{tab:frequencies_greigon} (low-frequency region) and~\ref{tab:frequencies_preigon} (high-frequency region). We carefully considered all remaining frequencies of both the low- and high-frequency regions in our search for linear combinations of the parent frequencies. We consider that a frequency is a combination (resp. a harmonic) frequency if the difference between the calculated combination (resp. harmonic) frequency and the candidate frequency is smaller than $f_{\rm res}$ \citep[e.g.][]{Zhang2018}. The third most dominant frequency in the Fourier spectrum is $f_{3}$ (= 0.721 d$^{-1}$) with an amplitude A$_{3}$ equal to 1.1 mmag. The uncertainties on all the significant frequencies are as large as $\epsilon_{f} = (0.04-3)\times 10^{-4}$ d$^{-1}$. We remark that it is the dominant one in the \Kepler SC periodogram \citep[see also][]{Uytterhoeven2011}. Since we also identified the first and second harmonics of $f_{3}$, namely $f_{34}$ (= 1.442 d$^{-1}$) and $f_{215}$ (= 2.163 d$^{-1}$), we consider that $f_{3}$ is a good candidate for the rotational frequency $f_{\rm rot}$ of the fast-rotating star in the system. Among the frequencies detected in both regions of the Fourier spectrum, we recognised the following three combination types: \\
\begin{enumerate}[(i)]
\item $f_{i}$= $nf_{\rm parent_{i}} \pm mf_{\rm parent_{j}}$ ($n,m$ =  1, 2, 3): \struutdown\\  That is to say linear combinations of the parent frequencies. Such frequencies are marked in blue on panels (a) of Fig.~\ref{fig:gmode-split} (for the low-frequency region: $f<5.0$ d$^{-1}$) and Fig.~\ref{fig:pmode_splt} (for the high-frequency region: $f\geq 5.0$ d$^{-1}$). The relevant information for the detected combinations is mentioned under the columns 'combs' and  'remark' in Tables~\ref{tab:frequencies_greigon} and~\ref{tab:frequencies_preigon}. An explanation of the used terminology is provided in Table~\ref{tab:terminology}. \struutdown
\item $f_{i}$ = $nf_{\rm parent} \pm m f_{\rm rot}$ ($n,m$ = 1, 2, 3):\struutdown\\ That is to say combinations of the parent frequencies (or their harmonics) and $f_{\rm rot}$ (or its harmonics). Such frequencies are marked in orange across the entire Fourier spectrum on panel (a) of Figs.~\ref{fig:gmode-split} and~\ref{fig:pmode_splt} (resp. $f<5.0$ d$^{-1}$ and $f\geq 5.0$ d$^{-1}$). We marked $f_{\rm rot}$ and its harmonics with red pluses on panel (b) of Fig.~\ref{fig:gmode-split}. These combinations are also listed in Tables~\ref{tab:frequencies_greigon} and~\ref{tab:frequencies_preigon}.\struutdown
\item $f_{i}$= $nf_{\rm parent} \pm mf_{\rm ach_{1,2}}$ ($n,m$ =  1, 2, 3):\struutdown\\ 
We furthermore detected two probable high-order harmonics of $f_{\rm ach}$ (see Section~\ref{sec:activity-photom}) in the Fourier spectrum: the first one ($f_{117} =0.0087$ d$^{-1}$) is called $f_{\rm ach1}$ and the second one ($f_{169} =0.0856$ d$^{-1}$) is called $f_{\rm ach2}$ in Table~\ref{tab:terminology}. These frequencies are marked in green in Tables~\ref{tab:frequencies_greigon} and ~\ref{tab:frequencies_preigon} as well as in panel (a) of Fig.~\ref{fig:gmode-split}. We detected a couple of combinations of the parent frequencies and both of these frequencies. \struutdown
\end{enumerate}
\section{The low-frequency region and rotational splitting of $g$ modes} \label{sec:low-freqs}
\begin{figure*}
\centering
	\includegraphics[width=17cm,keepaspectratio]{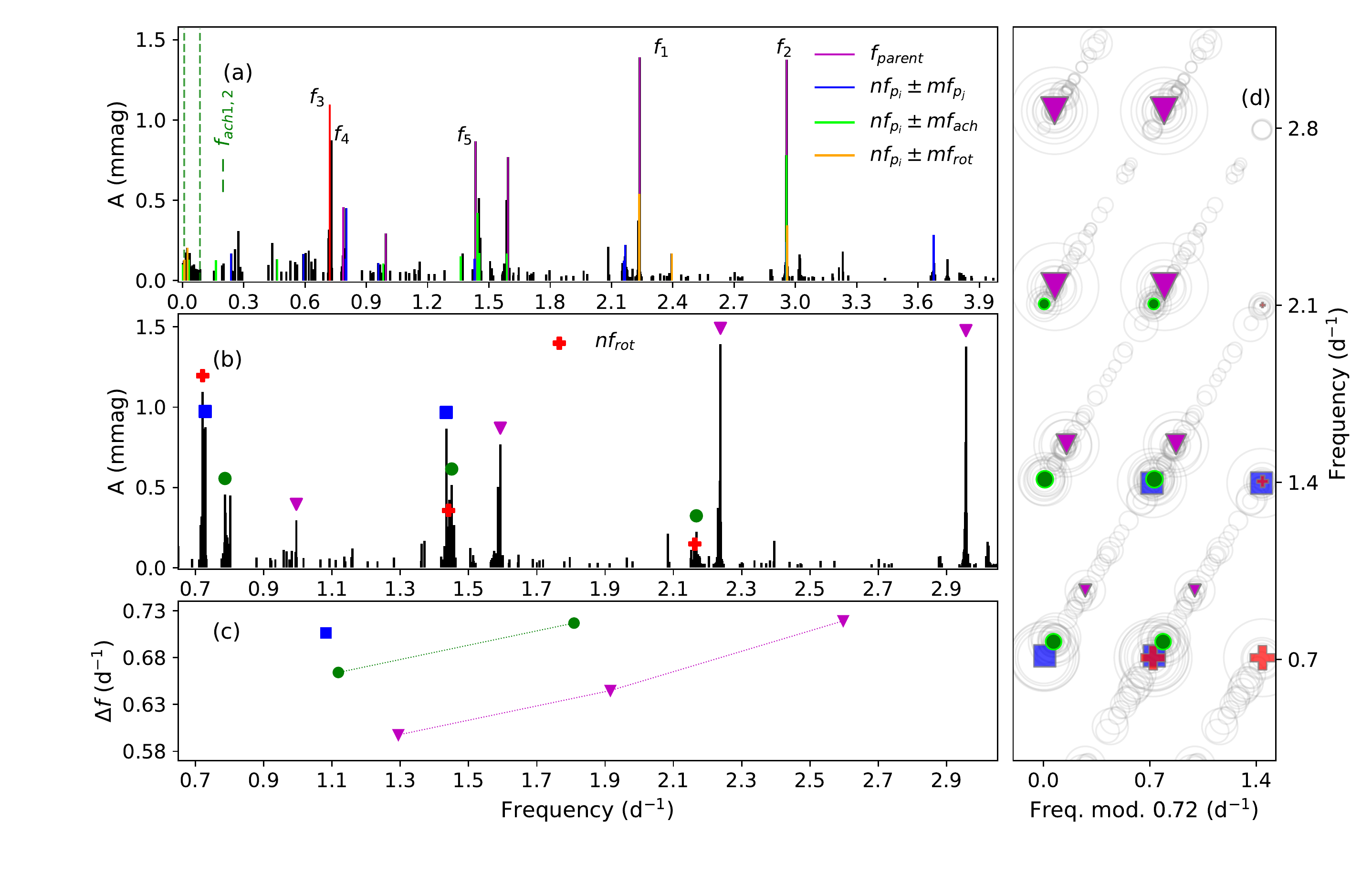}
	\caption{Low-frequency region ($f\leq 5$ d$^{-1}$) of the Fourier spectrum for the full \Kepler LC observations of KIC~6951642. (a) Different types of combination frequencies are detected among the significant frequencies of this region (Table~\ref{tab:frequencies_greigon}). For the terminology, see Table~\ref{tab:terminology}. $f_{1-5}$ represent the frequencies in descending order of amplitude (Table~\ref{tab:frequencies_greigon}.) (b) Identification of the rotational frequency: the frequencies shown with red pluses show the rotational frequency and its first and second harmonics (Table~\ref{tab:gmode_splt}). (c) Detection of rotationally split values for the multiplets in panel (b) (Table~\ref{tab:gmode_splt}). (d) The \'echelle diagram for the multiplets in panel (b). Grey circles represent the significant low frequencies. The marker's size is linked to the amplitude of the frequency. Colours and indicators follow the descriptions of panels (b) \& (c). }
	\label{fig:gmode-split}
\end{figure*}
\begin{table}
 \centering
	\caption{Multiplets of asymmetrically split $g$ modes.}
	\label{tab:gmode_splt}
	\begin{tabular}{lcccccl}
	\hline
	\hline
		$f_\mathrm{i}$& $f$    &  $A$  &  S/N  & $\Delta f\pm\sigma_\mathrm{\Delta f}$ & marker\\
	              &  d$^{-1}$  &  mmag &       &  d$^{-1}$                             &  \\
	\hline
    $f_{ 16}$ & 0.7866 &  0.4557 & 17 &                 & \textcolor{Green}{$\bullet$}\struutup
    \\
    $f_{ 13}$ & 1.4506 &  0.5148 & 22 & 0.664$\pm$0.004 & \textcolor{Green}{$\bullet$}            \\
    $f_{ 38}$ & 2.1673 &  0.2231 & 16 & 0.717$\pm$0.016 & \textcolor{Green}{$\bullet$}\struutdown \\
    $f_{ 27}$ & 0.9956 &  0.2943 & 13 &                 & \textcolor{Magenta}{$\blacktriangle$}\\
    $f_{ 8}$  & 1.5932 &  0.7688 & 29 & 0.598$\pm$0.029 & \textcolor{Magenta}{$\blacktriangle$}\\ 
    $f_{ 1}$  & 2.2380 &  1.3910 & 59 & 0.645$\pm$0.011 & \textcolor{Magenta}{$\blacktriangle$}\\ 
    $f_{ 2}$  & 2.9570 &  1.3764 & 77 & 0.719$\pm$0.017 & \textcolor{Magenta}{$\blacktriangle$}\\
    $f_{ 4}$  & 0.7290 &  0.8743 & 27 &                 & \textcolor{Blue}{$\blacksquare$}   \\
    $f_{ 5}$  & 1.4351 &  0.8659 & 30 & 0.706$\pm$0.012 & \textcolor{Blue}{$\blacksquare$}\struutdown\\
    \hline
	\end{tabular}
	\tablefoot{The mean spacing is $\Delta f_{\rm mean}$ = 0.675$\pm$0.044 d$^{-1}$. The error is the standard deviation associated with the mean spacing. The column 'marker' shows the symbols and colours for each multiplet of Fig.~\ref{fig:gmode-split}.}
\end{table}	
Panel (a) of Fig.~\ref{fig:gmode-split} presents a closer view of the low-frequency region of the object's Fourier spectrum ($f<5.0$ d$^{-1}$). We derived 309 frequencies with S/N $\geq$ 4.0. We list 97 of them with amplitudes larger than 0.08 mmag in Table~\ref{tab:frequencies_greigon}. Column $f_{i}$ of Table~\ref{tab:frequencies_greigon} shows the frequencies sorted in descending order of amplitude. The mean amplitude and mean S/N are A$_{\rm mean}$ = 0.121 mmag and S/N$_{\rm mean}$ = 8.4. The highest-amplitude frequency occurs at $f_{1}$ = 2.238$\pm$4$\times$10$^{-5}$ d$^{-1}$ (with A$_{1}$ = 1.3910$\pm$0.0015 mmag and S/N = 59).\\
Many $\gamma$ Dor/$\delta$ Sct stars experience intermediate to fast rotation. Hence, the first-order approximation in the traditional approximation of stellar rotation (TAR) does not hold anymore \citep[e.g.][]{Reese2008,Bouabid2013}. In this case, the frequency spacing is no longer a constant. When higher-order rotation terms become significant, we no longer expect symmetric multiplets \citep{Saio1981}. Some examples are mentioned in Table~\ref{tab:prev_study}. Our estimation of M$v$, based on distance derived for all Gaia DR2 targets by \citet{Bailer-Jones2018A} (Table~\ref{tab:photom_data}), is $\sim$0.92 mag for the entire system (ignoring the extinction as the star is located at $\sim$570 pc). By assuming a magnitude difference of at least 1.5 mag for the companion \citep{Batten1973}, we find M$v$ = 1.19 mag for the pulsating component. Accordingly, we derived $\log{\rm L/L_{\sun}} \approx 1.41$ from M$v$. Using T$_{\rm eff} = 7336$ K from Table~\ref{tab:atm_par}, we derive an average radius of 3.15 R$_{\sun}$. From the mass-luminosity relation $L = L_{\sun}(\frac{M}{M_{\sun}})^{3.9}$ \citep{Cox2000}, we estimate a mass of approximately 2.3 M$_{\sun}$ for the primary. Considering the critical Keplerian angular velocity $\Omega_{K}$ as $\Omega \ll \Omega_{K} \equiv \sqrt{GM/R^{3}}$; with G the universal constant of gravity, M the stellar mass and R the derived mean stellar radius and $f_{3}$ as the rotation frequency, we obtain $\Omega/\Omega_{K}$ = 75\% for the fast-rotating companion. Using the polar radius to derive $\Omega$ (R$_{\rm equatorial}$ = 3/2 R$_{\rm mean}$) results in an even larger $\Omega/\Omega_{K}$ ratio. Hence, we may expect to find multiplets that are affected by asymmetric splitting. \\
The second highest-amplitude frequency, $f_{2}$ = 2.957$\pm$4$\times$10$^{-5}$ d$^{-1}$, has an amplitude very similar to $f_{1}$, A$_{2}$ = 1.3764$\pm$0.0016 mmag and S/N = 77. We suggest that $f_{1}$ and $f_{2}$ are the members of a multiplet with $\Delta f$ values increasing from 0.598$\pm$0.029 d$^{-1}$ to 0.719$\pm$0.017 d$^{-1}$ (which lies close to $f_{3}$). We marked this multiplet with magenta triangles in panel (b) of Fig.~\ref{fig:gmode-split}. $f_{\rm rot}$ (Table~\ref{tab:terminology} \& Sect.~\ref{sec:fourier-comb}) and its harmonics are labelled with red plus marks. Panel (b) is a closer view of the region between 0.7 and 3.0 d$^{-1}$. In this region, we detected two other series of retrograde $g$ modes with a minimum spacing value of $\Delta f$ = 0.664$\pm$0.004 d$^{-1}$ and a maximum spacing value of $\Delta f$ = 0.717$\pm$0.016 d$^{-1}$ (which lies close to $f_{3}$). The quoted errors represent the standard deviation associated with the mean value of all three multiplets, that is to say $\Delta f\pm\sigma_{\Delta f}$ = 0.675$\pm$0.044 d$^{-1}$. Panel (c) of Fig.~\ref{fig:gmode-split} shows the $\Delta f$ values for the different $g$ modes, with their associated marker and colour in panel (b) and Table~\ref{tab:gmode_splt}. We suggest that the detected multiplets are due to intermediate or fast rotation, which causes them to be split asymmetrically. We marked these frequencies with the label 'RS' (Rotationally Split) in the column 'remark' of Table~\ref{tab:frequencies_greigon}. In panel (d) of Fig.~\ref{fig:gmode-split}, we plotted all the frequencies in an \'echelle diagram using the frequency modulo the (candidate) rotation frequency $f_{\rm rot}$ = 0.721 d$^{-1}$ on the X-axis. We see that the multiplets are arranged on wiggled vertical ridges, indicative of modes with the same degree $\ell$. However, we couldn't identify any period spacing pattern among $g$ modes (according to asymptotic relation for $g$ modes.) 

\section{The high-frequency region and rotational splitting of $p$ modes}\label{sec:high-freqs}
\begin{figure}
\centering
	\includegraphics[width=\columnwidth]{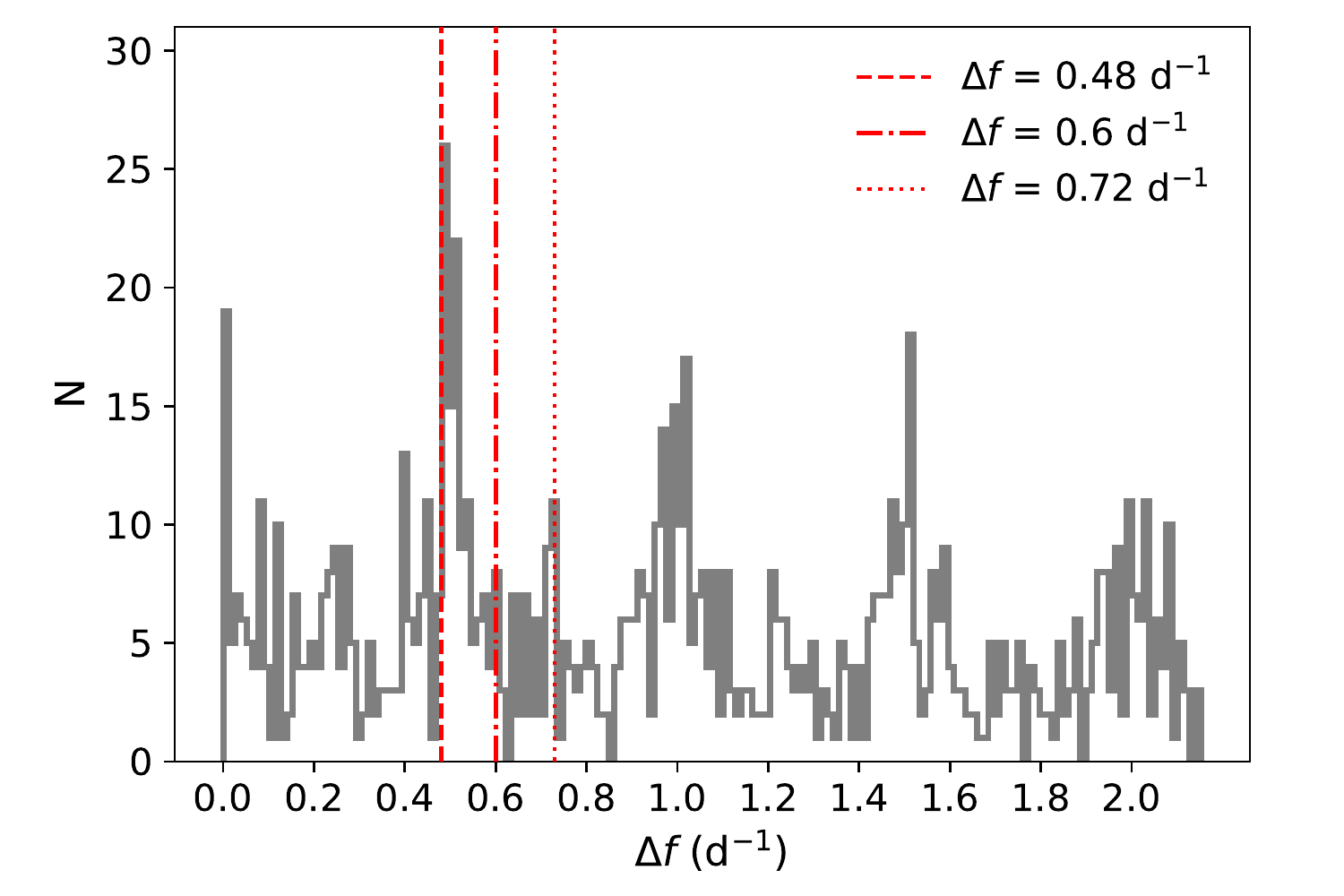}
	\caption{Histogram of all possible frequency spacings for the parent $p$ modes between 7.0-17.0 d$^{-1}$ (Table~\ref{tab:frequencies_preigon}). Three highest peaks, excluding the harmonics of 0.48, are indicated with red lines.}
	\label{fig:pmode_histo}
\end{figure}
\begin{figure*}
\centering
	\includegraphics[width=17cm]{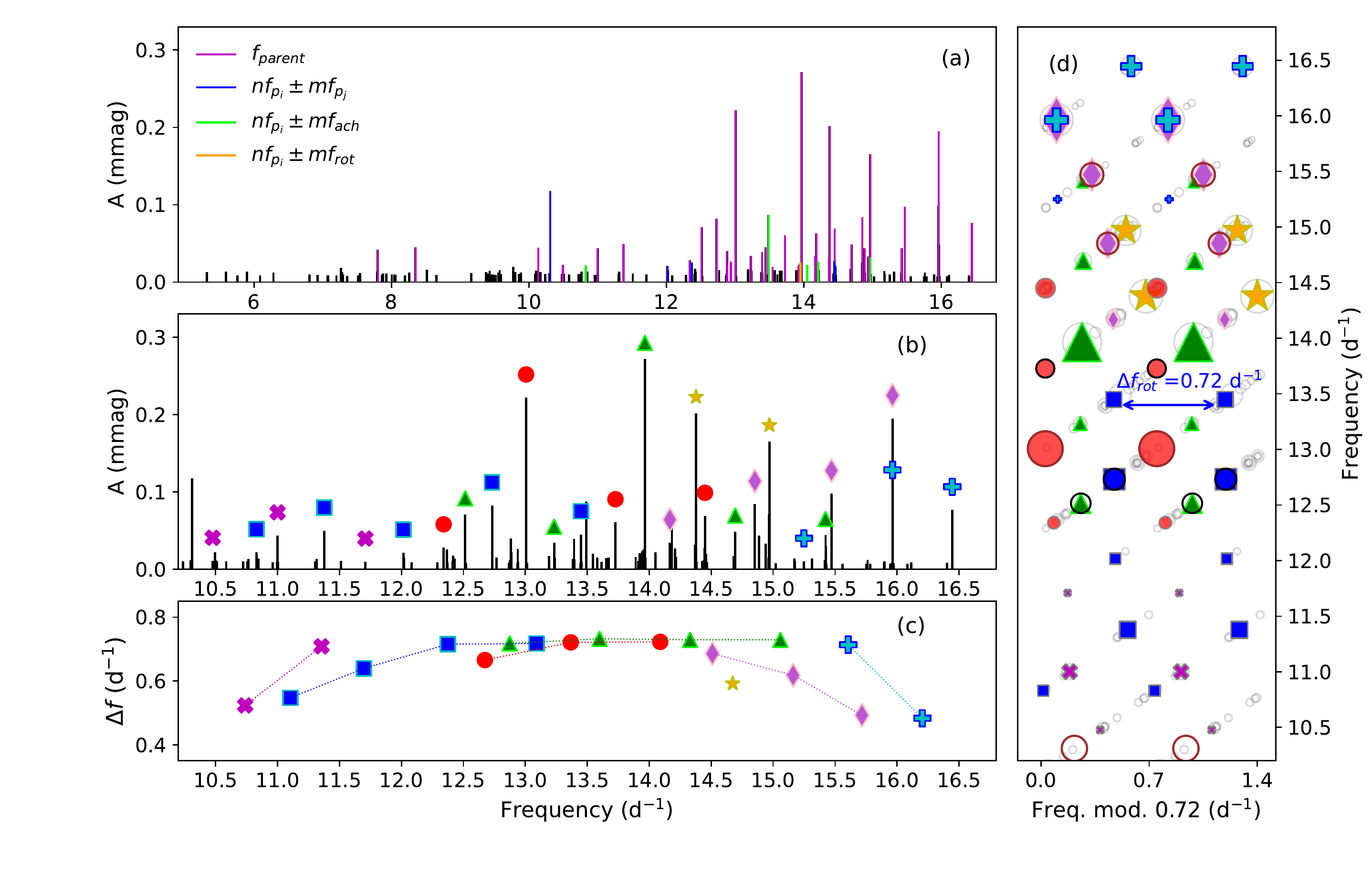}
	\caption{High-frequency region ($f\geq 5$ d$^{-1}$) of the Fourier spectrum for the full \Kepler LC observations of KIC~6951642. (a) Different types of combinations are detected among the significant frequencies (S/N$\geq$4)(Table~\ref{tab:frequencies_preigon}). (b) Detection of rotationally split $p$ modes in the interval (10.5-17) d$^{-1}$ ($\Delta f_{\rm rot} \equiv f_{3} = 0.721 $ d$^{-1}$)  (Table~\ref{tab:pmode_splt}).(c) Detection of rotationally split values for the multiplets in panel (b) (Table~\ref{tab:pmode_splt}). (d) The \'echelle diagram for the detected multiplets in panel (b). Grey circles: the significant high-frequencies. The marker's size is linked to the amplitude of the significant frequency. Black circles: The frequencies whose TDs show a similar long-term time delay trend (TDr in Table~\ref{tab:frequencies_preigon}). Brown circles: The frequencies showing incompatibility with general long-term time delay trend (TDir in Table~\ref{tab:frequencies_preigon}). Colours and indicators follow the descriptions of panels (b) \& (c).}
	\label{fig:pmode_splt}
\end{figure*}
\begin{table}
 \centering
	\caption{Multiplets of asymmetrically split $p$ modes.}
	\label{tab:pmode_splt}
	\begin{tabular}{lcccccl}
	\hline
	\hline
	$f_\mathrm{i}$& $f$    &  $A$  &  S/N  & $\Delta f\pm\sigma_\mathrm{\Delta f}$ & marker\\
	              &  d$^{-1}$  &  mmag &       &  d$^{-1}$                             &  \\
	\hline
    $f_{ 434}$ & 10.4774 &  0.0108 & 4  &                 & \textcolor{Magenta}{\bf{$X$}}\struutup\\
    $f_{ 230}$ & 11.0008 &  0.0436 & 15 & 0.523$\pm$0.050 & \textcolor{Magenta}{\bf{$X$}}\\
    $f_{ 463}$ & 11.7093 &  0.0098 & 4  & 0.708$\pm$0.020 & \textcolor{Magenta}{\bf{$X$}}\struutdown\\
    $f_{ 328}$ & 10.8299 &  0.0218 & 8  &                 & \textcolor{Blue}{$\blacksquare$}\\
    $f_{ 210}$ & 11.3772 &  0.0497 & 18 & 0.547$\pm$0.041 & \textcolor{Blue}{$\blacksquare$}\\
    $f_{ 331}$ & 12.0166 &  0.0213 & 8  & 0.639$\pm$0.006 & \textcolor{Blue}{$\blacksquare$}\\
    $f_{ 126}$ & 12.7318 &  0.0823 & 24 & 0.715$\pm$0.023 & \textcolor{Blue}{$\blacksquare$}\\
    $f_{ 223}$ & 13.4492 &  0.0450 & 15 & 0.717$\pm$0.023 & \textcolor{Blue}{$\blacksquare$}\struutdown\\
    $f_{ 288}$ & 12.3408 &  0.0281 & 11 &                 & \textcolor{Red}{$\bullet$}\\
    $f_{ 39}$  & 13.0064 &  0.2219 & 53 & 0.666$\pm$0.004 & \textcolor{Red}{$\bullet$}\\
    $f_{ 178}$ & 13.7274 &  0.0606 & 18 & 0.721$\pm$0.025 & \textcolor{Red}{$\bullet$}\\
    $f_{ 154}$ & 14.4500 &  0.0689 & 20 & 0.723$\pm$0.025 & \textcolor{Red}{$\bullet$}\struutdown\\
    $f_{ 146}$ & 12.5150 &  0.0708 & 22 &                 & \textcolor{Green}{$\blacktriangle$}\\
    $f_{ 260}$ & 13.2328 &  0.0341 & 12 & 0.718$\pm$0.024 & \textcolor{Green}{$\blacktriangle$}\\
    $f_{ 31}$  & 13.9651 &  0.2718 & 54 & 0.732$\pm$0.029 & \textcolor{Green}{$\blacktriangle$}\\
    $f_{ 213}$ & 14.6942 &  0.0485 & 16 & 0.729$\pm$0.028 & \textcolor{Green}{$\blacktriangle$}\\
    $f_{ 226}$ & 15.4231 &  0.0441 & 16 & 0.729$\pm$0.028 & \textcolor{Green}{$\blacktriangle$}\struutdown\\
    $f_{ 43}$  & 14.3778 &  0.2018 & 43 &                 & \textcolor{BurntOrange}{$\textbf{*}$}\\
    $f_{ 59}$  & 14.9697 &  0.1652 & 39 & 0.592$\pm$0.024 & \textcolor{BurntOrange}{$\textbf{*}$}\struutdown\\
    $f_{ 259}$ & 14.1668 &  0.0341 & 12 &                 & \textcolor{Orchid}{$\blacklozenge$}\\
    $f_{ 123}$ & 14.8528 &  0.0841 & 23 & 0.686$\pm$0.012 & \textcolor{Orchid}{$\blacklozenge$} \\
    $f_{ 107}$ & 15.4706 &  0.0976 & 28 & 0.618$\pm$0.014 & \textcolor{Orchid}{$\blacklozenge$} \\
    $f_{ 44}$  & 15.9634 &  0.1948 & 53 & 0.493$\pm$0.061 & \textcolor{Orchid}{$\blacklozenge$} \struutdown\\
    $f_{ 453}$ & 15.2490 &  0.0102 & 5  &                 & \textcolor{Cyan}{\textbf{$+$}}\\
    $f_{ 104}$ & 15.9627 &  0.0985 & 30 & 0.714$\pm$0.022 & \textcolor{Cyan}{\textbf{$+$}}\\
    $f_{ 135}$ & 16.4458 &  0.0766 & 27 & 0.483$\pm$0.065 & \textcolor{Cyan}{\textbf{$+$}}\struutdown\\
    \hline
	\end{tabular}
	\tablefoot{The mean spacing is $\Delta f_{\rm mean}$ = 0.665$\pm$0.084 d$^{-1}$. The error value is the standard deviation associated with the mean spacing. The column 'marker' shows the symbols and the colours of each multiplet in Fig.~\ref{fig:pmode_splt}.}
\end{table}
\begin{figure}
\centering
	\includegraphics[width=\columnwidth,keepaspectratio]{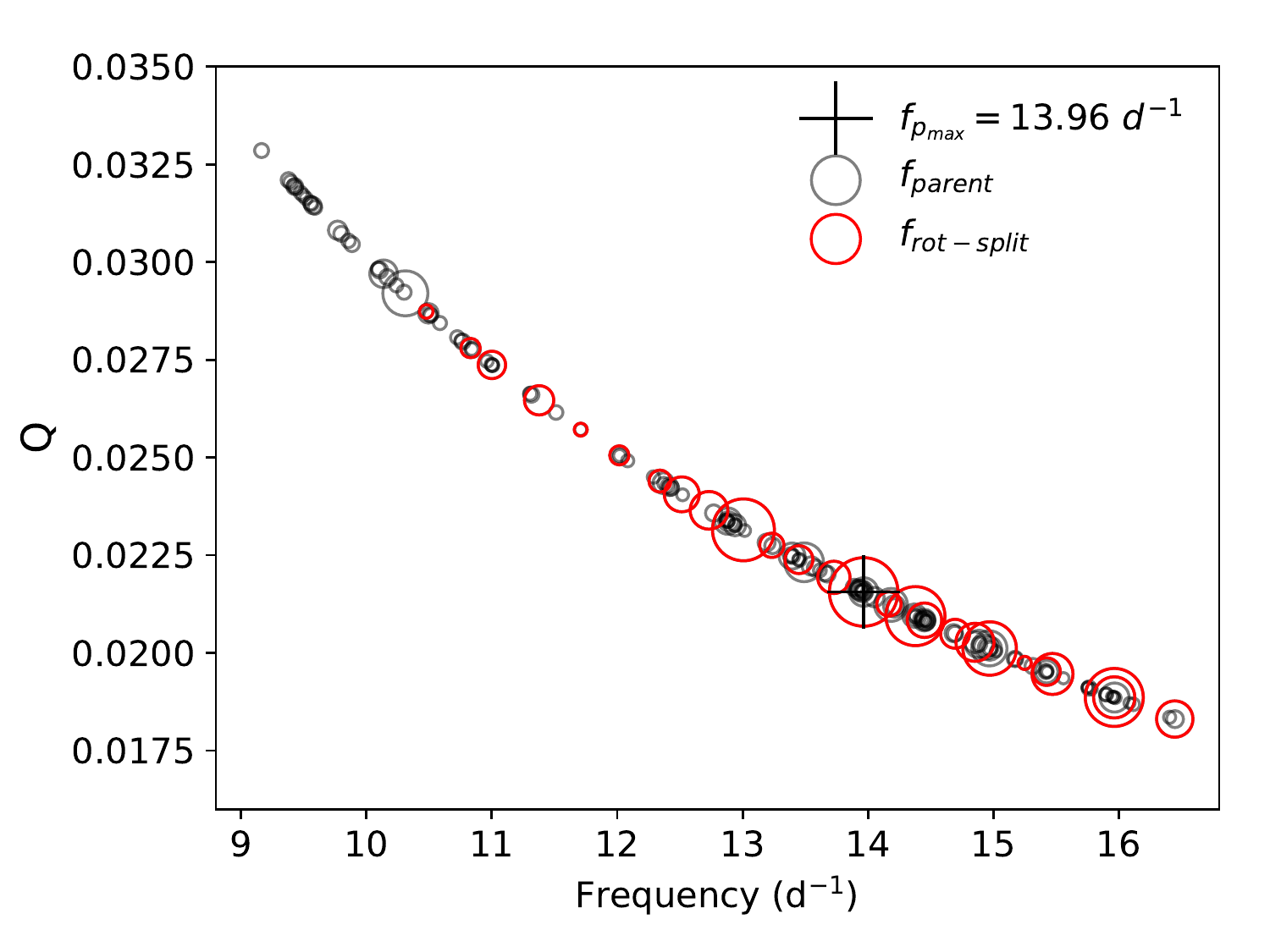}
	\caption{Pulsation constants Q for the high frequencies revealed by the Fourier spectrum of KIC~6951642. Black circles show the parent frequencies (see Table~\ref{tab:frequencies_preigon}). Red circles show the rotationally split $p$ modes (see Table~\ref{tab:pmode_splt}). The marker's size is associated with the amplitude of the $p$ modes.}
	\label{fig:pmode_Q}
\end{figure}
Panel (a) of Fig.~\ref{fig:pmode_splt} presents a closer view of the high-frequency region of the Fourier spectrum ($f\geq5.0$ d$^{-1}$). We derived 284 frequencies with S/N$\geq$ 4.0. We list 55 of them with amplitudes larger than 0.019 mmag in Table~\ref{tab:frequencies_preigon}. Column $f_{i}$ of Table~\ref{tab:frequencies_preigon} shows the frequencies sorted in descending order of amplitude. The mean amplitude and mean S/N are A$_{\rm mean}$ = 0.0193 mmag and S/N$_{\rm mean}$ = 7.72. The highest-amplitude frequency occurs at $f_{31}$ = 13.9651 $\pm$ 2$\times$10$^{-5}$ d$^{-1}$ (with A$_{31}$ = 0.2718$\pm$0.008 mmag and S/N = 54).\\
In the search for regular patterns, we calculated all possible frequency spacings between all the pairs of parent frequencies ($p$ modes). These parent frequencies are labelled as $f_{\rm parent}$ in magenta under the column 'remark' in Table~\ref{tab:frequencies_preigon} (Sect.~\ref{sec:fourier-comb}). We illustrate the distribution of the spacings in Fig.~\ref{fig:pmode_histo}. The most frequent spacing values are located at $\sim$0.48 d$^{-1}$ and its harmonics. Excluding the mentioned most frequent spacings the other two peaks are located at 0.6 d$^{-1}$, and 0.72 d$^{-1}$. We recall that 0.721 d$^{-1}$ equals the rotation frequency as discussed in Sect.~\ref{sec:fourier-comb}. The first detection represents a quintuplet of rotationally split $p$ modes centred around $f_{31}$, with mean frequency spacing $\Delta f$ = 0.727$\pm$0.005 d$^{-1}$ (equivalent to $f_{\rm rot}$). In panel (b) of Fig.~\ref{fig:pmode_splt}, we marked the frequencies of this multiplet with green triangles. The same figure illustrates six more rotationally split multiplets of retrograde and prograde $p$ modes. Panel (c) of Fig.~\ref{fig:pmode_splt} displays the $\Delta f$ values of these multiplets. We list these values in Table~\ref{tab:pmode_splt}. The uncertainty in the frequency spacing, $\sigma_{\Delta f}$, represents the standard deviation of the mean value. The detected splittings are asymmetric on both sides. Their values range from 0.483$\pm$0.065 d$^{-1}$ to 0.729$\pm$0.028 d$^{-1}$ (similar to $f_{3}$). The mean spacing for all seven multiplets is $\Delta f_{\rm mean}$ = 0.665$\pm$0.084 d$^{-1}$ d$^{-1}$. We suggest that all detected multiplets are due to intermediate or fast rotation, which causes them to be split asymmetrically. Thus, we observe the same behaviour as with the multiplets of the $g$ modes. In Table~\ref{tab:pmode_splt}, we list the frequency, the amplitude and the S/N of each multiplet. We indicate these frequencies with the label 'RS' (Rotationally Split) in the column 'remark' of Table~\ref{tab:frequencies_greigon}.\\
In panel (d) of Fig.~\ref{fig:pmode_splt}, we plotted all the frequencies between 10-17 d$^{-1}$ in an \'echelle diagram using the frequency modulo the rotation frequency $f_{\rm rot}$ = 0.721 d$^{-1}$ on the X-axis. We present the frequencies with grey circles with their sizes associated with their amplitudes. The frequencies whose TDs show a similar long-term trend (annotated as 'TDr' in Table~\ref{tab:frequencies_preigon} and in the upper panel of Fig.~\ref{fig:tds46fs}) are presented with black circles in Fig.~\ref{fig:pmode_splt}, six ($f_{~39,~107,~123,~126,~146,~178}$) of which belong to rotationally split multiplets. The frequencies showing incompatibility with the general trend ('TDir' in Table~\ref{tab:frequencies_preigon} and in the bottom panel of Fig.~\ref{fig:tds46fs}) are represented by brown circles, three ($f_{31,~43,~44}$) of which belong to multiplets ($f_{31}\equiv f_{\rm p_{\rm max}}$). \\
The pulsation constants \citep{Breger1990} Q of the parent (shown as black circles in Fig.~\ref{fig:pmode_Q}) and the rotationally split $p$ modes (shown as red circles in Fig.~\ref{fig:pmode_Q}) were derived using T$_{\rm eff}$ and $\log{g}$ from Sect.~\ref{sec:orb-new} as well as M$_{\rm bol}$ based on the Gaia DR3 parallax \citep{Gaia2021} (see Table~\ref{tab:photom_data}). 
\section{Discussion and conclusions}\label{sec:conclusion}
KIC~6951642 is a \Kepler object of magnitude Kp = 9.70 mag. It was introduced as a candidate hybrid $\gamma$ Dor\,--\,$\delta$ Sct star from studies of the \Kepler light curves by \citet{Uytterhoeven2011}, who reported the highest-amplitude mode of 0.721 d$^{-1}$, and by \citet{Fox-Machado2017}. The low-amplitude variability in the RVs collected from multi-epoch spectroscopy with \textsc{HERMES} \citep{Lampens2018} as well as the study of the TDs of nine high frequencies \citep{Murphy2018} suggest that KIC~6951642 is a (single-lined) binary system. Spectroscopic analysis shows that it is a normal F0-type star with T$_{\rm eff}$ = 7336$\pm$186 K. The combination of the RV curve with the TDs of eight significant high frequencies detected in the \Kepler LC Fourier spectrum lead to an updated orbit with $P_{\rm orb}$ almost equal to 1800~d. Due to the incompleteness of the data (e.g. RV data are lacking in the phase intervals 0-0.2 and 0.8-0.95, cf. Fig.~\ref{fig:combined_solution}) and the lack of consistent behaviour in the TDs, the confirmation of its binary nature is not (yet) feasible and more spectroscopic observations will be needed. Currently, we consider it as a \textit{possible} binary system.\\
Although the spectropolarimetric observations of KIC~6951642 do not show any Zeeman signature as evidence of a magnetic field, we detected a sinusoidal behaviour with a period of $\sim$3.2 yr from analysis of S$_{\rm ph}$ in the \Kepler LC light curve ($f_{\rm ac}$ = 0.00085$\pm$0.0002 d$^{-1}$). Accordingly, at the first date of the spectropolarimetric observations (Apr. 19, 2019) the star was in a state of minimum activity (in the 1.96$^{\rm th}$ year of the \Kepler mission, see panel (b) of Fig.~\ref{fig:magnetic_cycle}). This could be the reason for the unsuccessful detection of magnetic activity from the spectropolarimetric observations. The same argument explains why the TESS light curve has a lower amplitude compared to the \Kepler LC light curve. The first date of the TESS observations (Aug. 15, 2019) occurred during a state of minimum activity (in the 2.29$^{\rm th}$ year of the \Kepler mission, see panel (b) of Fig.~\ref{fig:magnetic_cycle}). Unfortunately, since the cycle length is similar to the time span of the observations, we do not know what causes the suggested activity.\\ 
The atmospheric stellar parameters locate KIC~6951642 in the overlapping area of the $\delta$ Sct\,--\,$\gamma$ Dor instability strips. We used the relation R\,$\sin{i} = {v\sin{i}\,.\,P_{\rm rot}}\,/\,50.58$  to infer a lower limit for the stellar radius, with $v$ the equatorial velocity, $v\sin{i}$ the projected rotational velocity (\kms), and P$_{\rm rot}$ the estimated rotation period (d). This gives a minimum radius of 3.4 R$_{\sun}$, that is to say too large for a normal F0-type star. On the other hand, the study of $\delta$ Sct stars in eclipsing binaries \citep{Garcia2017} demonstrates that somewhat evolved $\delta$ Sct stars have radii larger than normal. The radius derived from the Gaia DR2 distance \citep[Table~\ref{tab:photom_data}][]{Bailer-Jones2018A}, and T$_{\rm eff}$ (cf. Table~\ref{tab:atm_par}) equals 3.15 R$_{\sun}$, almost the value derived by assuming $f{_3}$ = $f_{\rm rot}$. Hence, this confirms that the fast-rotating object in KIC~6951642 is an evolved F0-type star. \\
From the Fourier spectrum of the \Kepler LC light curve, 593 significant frequencies with S/N$\geq 4$ were extracted. The spectrum also reveals a strong regularity in both the low- and high-frequency regions. In the high-frequency region, we detected several $p$ modes in the range of 10-17 d$^{-1}$. The $p$ mode of highest amplitude is $f_{p_{\rm max}}$ = 13.965 d$^{-1}$ (A$_{\rm p_{\rm max}} =$ 0.27 mmag). We detected seven rotationally split multiplets comprising 26 $p$ modes with $\Delta f_{\rm mean} $ = 0.665$\pm$0.084 d$^{-1}$. \\
In the low-frequency region, we identified $f_{3}$ = 0.721 d$^{-1}$ and its first and second harmonics. We detected two frequencies close to high harmonics of $f_{\rm ac}$, a probable signature of stellar activity, in the Fourier spectrum ($f_{\rm ach1}$) $f_{177}$ = 0.0085 d$^{-1}$ and $f_{169}$= 0.0856 d$^{-1}$ ($f_{\rm ach2}$). We also detected several frequencies that are linear combinations of one of these three frequencies ($f_{\rm rot}$, $f_{\rm ach1}$ and $f_{\rm ach2}$) or their second and third harmonics with either a $g$  or a $p$ mode. We cannot explain the occurrence of such high harmonics of $f_{\rm ac}$ (i.e. 10$f_{\rm ac}$ and 100$f_{\rm ac}$). We identified the two most dominant frequencies in the Fourier spectrum ($f_{1}$ = 2.238 d$^{-1}$ and $f_{2}$ = 2.957 d$^{-1}$, A$_{1,2}\approx$ 1.4 mmag) as members of a retrograde multiplet split by rotation. We detected two other multiplets of rotationally split $g$ modes. The mean frequency spacing for the three multiplets of $g$ modes equals $\Delta f_{mean}$ = 0.675$\pm$0.044 d$^{-1}$. Accordingly, we associate the multiplets of $g$  and $p$ modes to the fast-rotating (component of) KIC~6951642 ($\Omega/\Omega_{\rm K}$ = 75\%), suggesting that $f_{3}$ is the rotational frequency. From the detected $\Delta f_{\rm mean}$ for both $g$  and $p$ modes, we derive a ratio of 1.02 $\pm$ 0.14 for the core-to-surface rotation, that is to say compatible with solid-body rotation. We couldn't identify any sensible period-spacing pattern matching the usual properties of $\gamma$ Dor pulsations. \\
In summary, this study confirms the occurrence of $\gamma$ Dor pulsation modes covering an extended range from 0.72 to 2.4 d$^{-1}$ due to significant stellar rotation and $\delta$ Sct pulsation modes covering the frequency range 10-17 d$^{-1}$ for the fast-rotating and (probably) active (companion) star KIC~6951642. Hence, we confirm the genuine hybrid pulsating nature of this object. \\
The binarity of KIC~6951642 is not yet certain, although a combined (RV+TD) solution is partly feasible. The reason why some $p$ modes show a similar long-term behaviour of their TDs while others (such as the strongest $p$ mode) do not, is an obstacle to this interpretation. However, it could be that some of the modes show intrinsic phase variability. \\
The next step would be to perform an evolutionary plus seismic modelling of this hybrid pulsator, which could become challenging due to the combined effects of significant rotation and probable stellar activity, as well as possible binarity. 
\begin{acknowledgements}
ASG acknowledges support from the Max Planck Society grant 'Preparations for PLATO science'. The authors appreciate the discussions with Dr. Daniel Reese at LESIA, Observatoire de Paris. The authors thank the \Kepler and TESS teams for their efforts to provide all the observational data and light curves to the public and the referee for constructive comments. The authors also thank the \textsc{HERMES} Consortium for enabling the production of the high-resolution ground-based spectra. This research made use of LightKurve, a Python package for \Kepler and TESS data analysis \citep{Lightkurve2018}. This work is based on data from the European Space Agency (ESA) mission {\it Gaia} (\url{https://www.cosmos.esa.int/gaia}), processed by the {\it Gaia} Data Processing and Analysis Consortium (DPAC,\url{https://www.cosmos.esa.int/web/gaia/dpac/consortium}). Funding for the DPAC has been provided by national institutions, in particular the institutions participating in the {\it Gaia} Multilateral Agreement.
\end{acknowledgements}
\bibliographystyle{aa}
\bibliography{Samadi.bib}
\begin{appendix}
\onecolumn
\setcounter{table}{0}
\renewcommand{\thetable}{A\arabic{table}}
\setcounter{figure}{0}
\renewcommand{\thefigure}{A\arabic{figure}}
\begin{table}
\centering
\caption{Journal of the \textsc{HERMES} spectra of KIC~6951642.} 
\begin{tabular}{crcrr}
\hline\noalign{\smallskip}
     BJD    &    Exp    &    S/N    & RV & e\_RV \\   
\hline\noalign{\smallskip}
2455820.493929  &   1800  &    78.14 &     -4.79    &	4.22\\
2455820.515356  &   1800  &    77.46 &     -5.13    &	4.68\\
2455823.517590  &   1500  &    62.50 &     -5.37    &	3.83\\
2455823.535549  &   1500  &    64.82 &     -4.89    &	4.13\\
2456515.563844  &   1000  &    47.08 &     -0.79    &	3.40\\
2456515.575998  &   1000  &    49.68 &     -1.59    &	3.19\\
2456601.325355  &    600  &    41.73 &     -1.40    &	4.04\\
2456605.334439  &    600  &    ---    &     0.20    &	3.57\\
2456605.341963  &    600  &    ---    &    -1.62    &	4.45\\
2457512.668146  &   1400  &    56.63  &    -3.24    &	3.19\\
2457514.527796  &   1700  &    56.23  &    -2.83    &	2.87\\
2457915.617435  &   1000  &    46.20  &    -10.86   &	4.64\\
2458679.501538  &    900  &    60.62  &      1.24   &	4.53\\
2458780.449374  &   1000  &    53.29  &     -0.01   &	3.42\\
2458787.333903  &    720  &    55.26  &      0.11   &	4.25\\
2458997.505925  &   1000  &    60.48  &     -0.43   &	3.19\\
2459091.424074  &   1000  &    51.57  &     -1.80   &	4.72\\
2459110.497660  &   1800  &    39.23  &     -1.99   &	2.49\\
2459391.482595  &   1000  &    45.77  &     -5.36   &	3.24\\
2459498.335626  &   1000  &    50.17  &     -8.16   &   3.50\\ \noalign{\smallskip}\hline\noalign{\smallskip}
\end{tabular}
\tablefoot{Barycentric Julian dates (BJD), exposure times (Exp), signal-to-noise ratio's (S/N's), radial velocities (RV) and their corresponding uncertainties (e\_RV).}
\label{tab:journal}
\end{table}
\begin{figure*}
\centering
	\includegraphics[height=7cm,keepaspectratio]{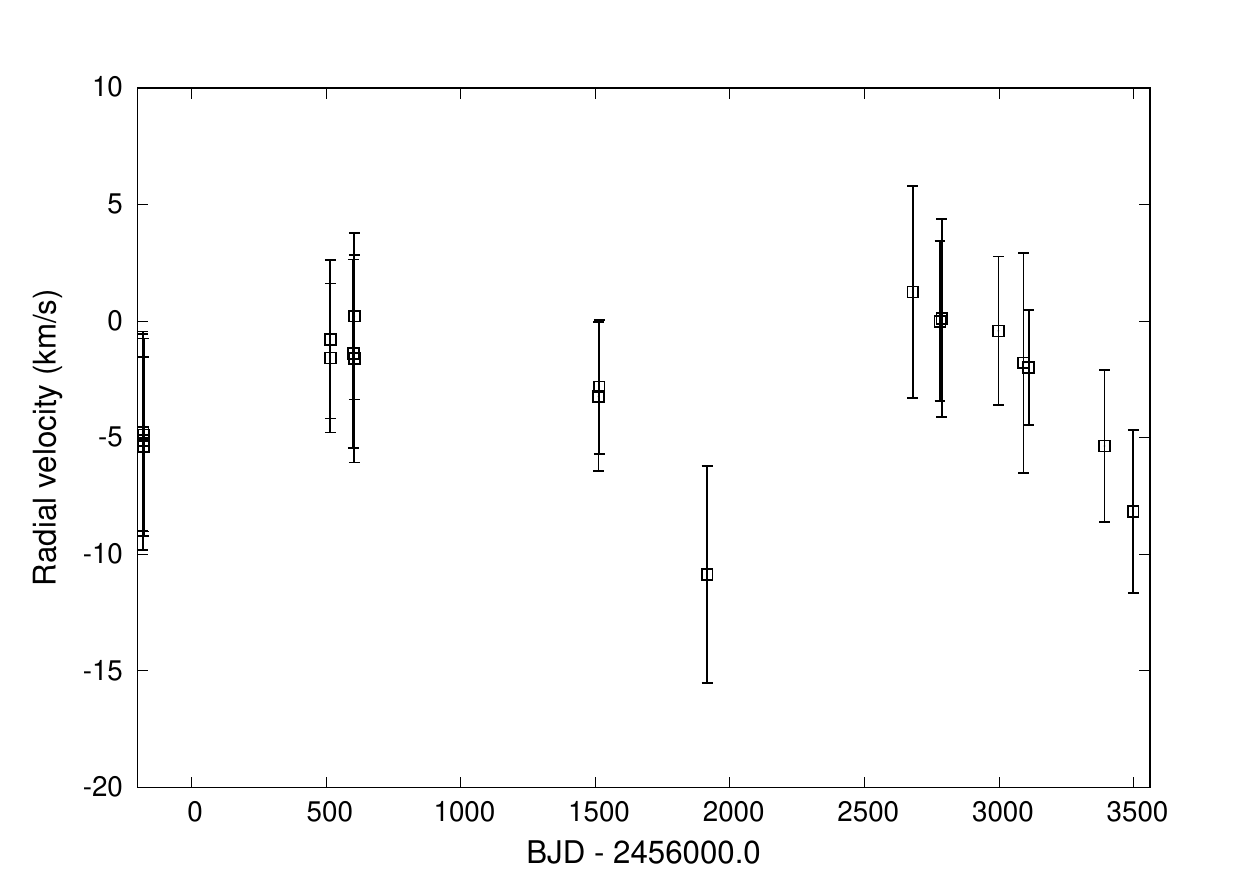} 
	\caption{RVs from the \textsc{HERMES} spectra of KIC~6951642 acquired over several years.}
	\label{fig:rvs}
\end{figure*}
\begin{figure*}
\centering
	\includegraphics[width=17cm, height=10cm,keepaspectratio]{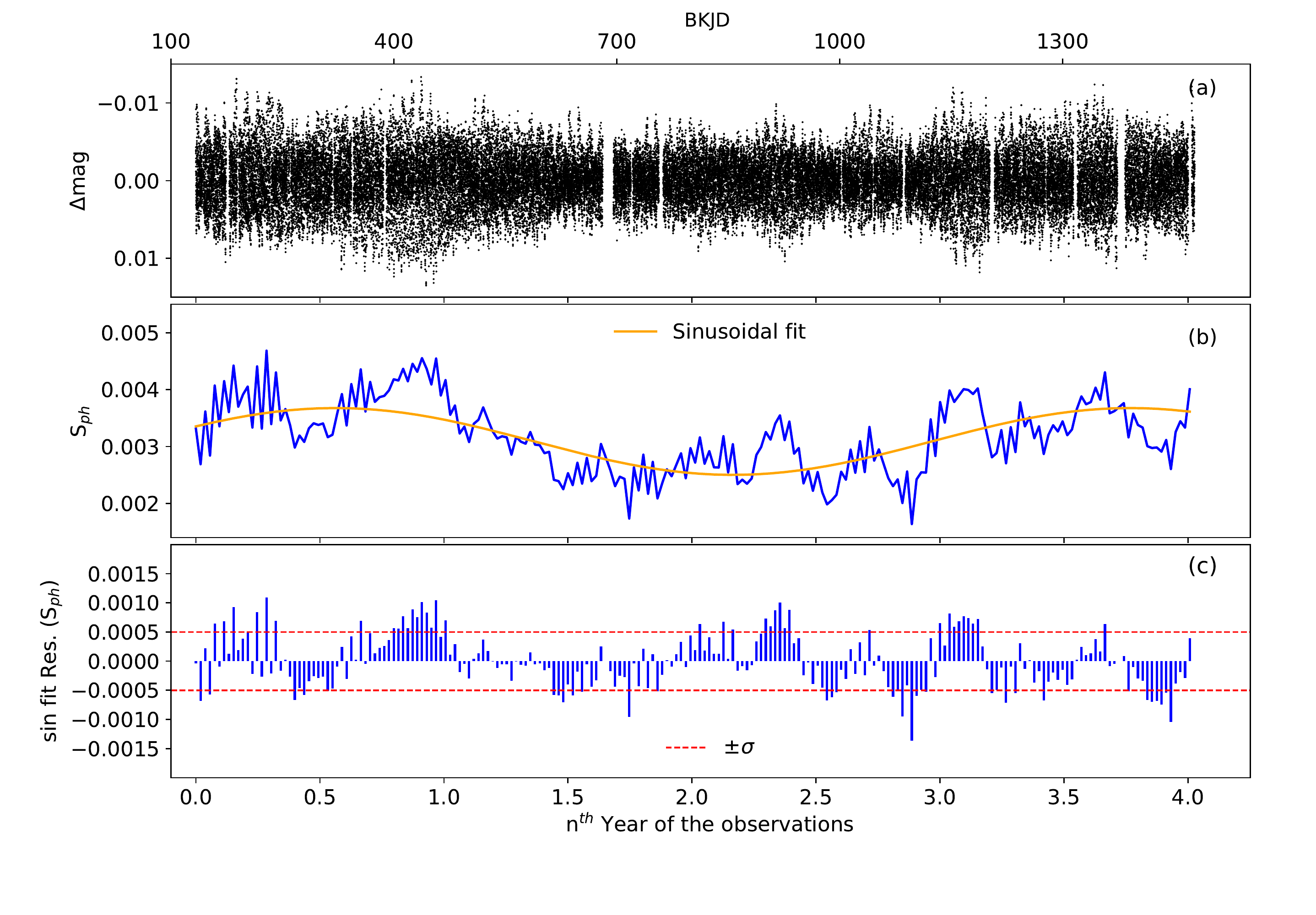}
	\caption{The photospheric magnetic activity proxy S$_{\rm ph}$ of KIC~6951642. (a) \Kepler LC light curve (1470.46 d) (b) Dispersion of the light curve on time scales of 5$f_{\rm rot}$ ($f_{\rm rot}$ =  0.721 d$^{-1}$). Red dashed lines are an illustration of boundaries of maximum magnetic and minimum activity cycles of star. Orange curve presents the sinusoidal fit to S$_{\rm ph}$ in different cycles. }
	\label{fig:magnetic_cycle}
\end{figure*}
\begin{figure*}
\centering
	\includegraphics[height=8cm,keepaspectratio]{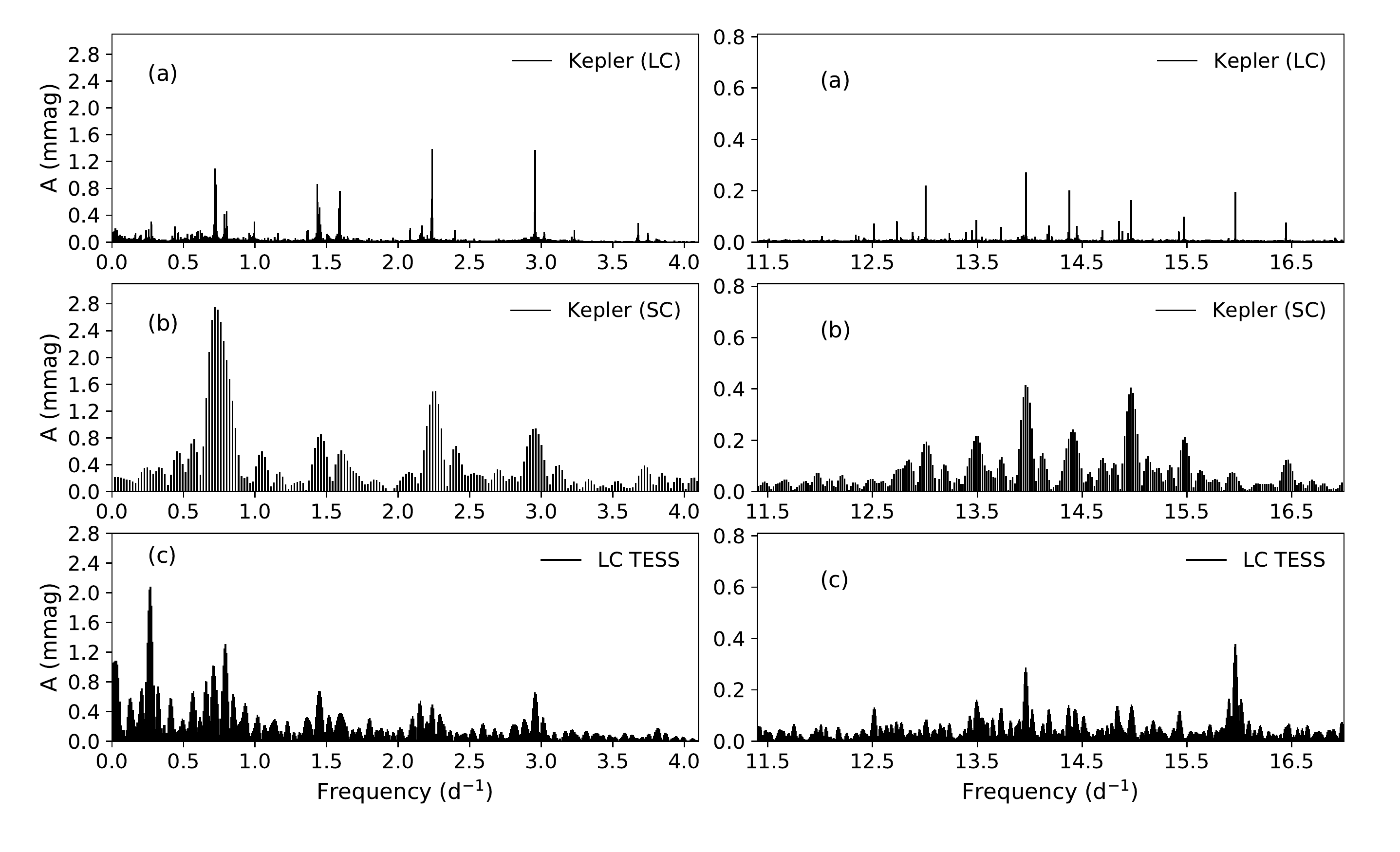}
	\caption{ A close-view to low- and high-frequency regions in Fourier Spectra KIC~6951642 from (a) \Kepler Long Cadence (b) \Kepler Short Cadence (c) TESS 1800 sec, sector 15. (Table~\ref{tab:LC_info}).}
	\label{fig:pergrams_zoom}
\end{figure*}

\begin{longtable}{clcccl}
\caption{Most dominant frequencies in low-frequency region (Fig.~\ref{fig:gmode-split}).}\\
\hline\hline
$f_{\mathrm i}$ & $f$ (d$^{-1}$) \footnotemark& A (mmag) \footnotemark  & S/N & comb.\footnotemark  & remark\struutup\struutdown\\    
\hline 
\endfirsthead
\caption{continued.}\\
\hline\hline
$f_{\mathrm i}$ & $f$ (d$^{-1}$) & A (mmag)   & S/N & comb. & remark\struutup\struutdown\\  
\hline
\endhead
\hline
\endfoot
\label{tab:frequencies_greigon}
$f_{ 1}$  &  2.2380  &  1.3910  &  59 & \textcolor{magenta}{$f_{\rm parent}$}                                                & RS  \struutup\\
$f_{ 2}$  &  2.9570  &  1.3764  &  77 & \textcolor{magenta}{$f_{\rm parent}$}                                                & RS  \\
$f_{ 3}$  &  0.7210  &  1.0962  &  31 & \textcolor{red}{$f_{\rm rot}$}                                                       &   \\
$f_{ 4}$  &  0.7290  &  0.8743  &  27 & \textcolor{red}{$f_{\rm rot}$}+\textcolor{PineGreen}{$ f_{\rm ach1}$}                & RS \\
$f_{ 5}$  &  1.4351  &  0.8659  &  30 & \textcolor{magenta}{$f_{\rm parent}$}                                                & RS \\
$f_{ 6}$  &  0.7249  &  0.8654  &  27 &                                                                                      &   \\
$f_{ 7}$  &  2.9559  &  0.7815  &  52 & \textcolor{PineGreen}{$ f_{\rm ach1}$}$+2f_{ 5}$                                     &   \\
$f_{ 8}$  &  1.5932  &  0.7688  &  29 & \textcolor{magenta}{$f_{\rm parent}$}                                                & RS  \\
$f_{ 9}$  &  0.7215  &  0.7507  &  24 &                                                                                      &   \\
$f_{ 10}$ &  0.7258  &  0.6184  &  21 &                                                                                      &   \\ 
$f_{ 11}$ &  2.9576  &  0.5642  &  40 &                                                                                      &   \\
$f_{ 12}$ &  2.2360  &  0.5413  &  28 & $f_{ 2} - $\textcolor{red}{$f_{\rm rot}$}                                            &   \\
$f_{ 13}$ &  1.4506  &  0.5148  &  22 & \textcolor{PineGreen}{$ f_{\rm ach1}$}+2\textcolor{red}{$f_{\rm rot}$}               & RS  \\
$f_{ 14}$ &  1.5857  &  0.5029  &  21 &                                                                                      &   \\
$f_{ 15}$ &  0.7224  &  0.4926  &  17 &                                                                                      &   \\
$f_{ 16}$ &  0.7866  &  0.4557  &  17 & \textcolor{magenta}{$f_{\rm parent}$}                                                & RS \\
$f_{ 17}$ &  1.5927  &  0.4550  &  21 &                                                                                      &   \\
$f_{ 18}$ &  0.8020  &  0.4499  &  17 & $f_{ 1} - f_{ 5}$                                                                    &   \\
$f_{ 19}$ &  1.4436  &  0.4224  &  19 & \textcolor{PineGreen}{$ f_{\rm ach1}$}+$f_{ 5}$                                      &   \\
$f_{ 20}$ &  2.2305  &  0.3730  &  22 &                                                                                      &   \\
$f_{ 21}$ &  2.9546  &  0.3453  &  27 &                                                                                      &   \\
$f_{ 22}$ &  2.9592  &  0.3433  &  28 & \textcolor{red}{$f_{\rm rot}$}+$f_{ 1}$                                              &   \\
$f_{ 23}$ &  0.7885  &  0.3418  &  13 &                                                                                      &   \\
$f_{ 24}$ &  0.7175  &  0.3180  &  12 &                                                                                      &   \\
$f_{ 25}$ &  0.2746  &  0.3091  &  12 & 2\textcolor{PineGreen}{$ f_{\rm ach1}$} + 3\textcolor{PineGreen}{$ f_{\rm ach2}$}    &   \\
$f_{ 26}$ &  1.4461  &  0.3072  &  15 &                                                                                      &   \\
$f_{ 27}$ &  0.9956  &  0.2943  &  13 & \textcolor{magenta}{$f_{\rm parent}$}                                                & RS \\
$f_{ 28}$ &  2.2388  &  0.2854  &  19 &                                                                                      &   \\
$f_{ 29}$ &  3.6768  &  0.2847  &  32 &  $2f_{ 2}-f_{ 1}$                                                                    &   \\
$f_{ 30}$ &  1.4498  &  0.2843  &  14 & \textcolor{PineGreen}{$ f_{\rm ach1}$}+2\textcolor{red}{$f_{\rm rot}$}               &   \\
$f_{ 32}$ &  1.4581  &  0.2660  &  14 & 2$f_{4}$                                                                             &   \\
$f_{ 33}$ &  0.7156  &  0.2647  &  10 &                                                                                      &   \\
$f_{ 34}$ &  1.4418  &  0.2562  &  13 & 2\textcolor{red}{$f_{\rm rot}$}                                                      &   \\
$f_{ 35}$ &  2.9539  &  0.2411  &  22 &                                                                                      &   \\
$f_{ 36}$ &  0.4395  &  0.2327  &  10 &  $f_{ 5}-f_{ 27}$                                                                    &   \\
$f_{ 37}$ &  1.4540  &  0.2235  &  12 &                                                                                      &   \\
$f_{ 38}$ &  2.1673  &  0.2231  &  16 &                                                                                      &   \\
$f_{ 40}$ &  2.0837  &  0.2111  &  14 & $2f_{ 5}-f_{ 16}$                                                                    &   \\
$f_{ 42}$ &  0.7913  &  0.2022  &  9  &                                                                                      &   \\
$f_{ 45}$ &  0.2575  &  0.1939  &  8  & 3\textcolor{PineGreen}{$f_{\rm ach2}$}                                               &   \\
$f_{ 46}$ &  0.7943  &  0.1896  &  8  & 3\textcolor{red}{$f_{\rm rot}$} + $f_{ 2}$                                           &   \\
$f_{ 47}$ &  0.6183  &  0.1877  &  8  &                                                                                      &   \\
$f_{ 48}$ &  3.2316  &  0.1815  &  21 &                                                                                      &   \\
$f_{ 50}$ &  0.6023  &  0.1723  &  7  &                                                                                      &   \\
$f_{ 52}$ &  1.4531  &  0.1707  &  10 &                                                                                      &   \\
$f_{ 53}$ &  0.2378  &  0.1694  &  7  &                                                                                      &   \\
$f_{ 54}$ &  1.5878  &  0.1693  &  10 &                                                                                      &   \\
$f_{ 55}$ &  1.4359  &  0.1675  &  10 &                                                                                      &   \\
$f_{ 56}$ &  1.3714  &  0.1675  &  10 &                                                                                      &   \\
$f_{ 57}$ &  2.3951  &  0.1671  &  13 &                                                                                      &   \\
$f_{ 58}$ &  0.5921  &  0.1662  &  7  &                                                                                      &   \\
$f_{ 60}$ &  3.0202  &  0.1613  &  16 &  $f_{ 5}$+2\textcolor{PineGreen}{$ f_{\rm ach1}$}                                    &   \\
$f_{ 62}$ &  0.7196  &  0.1587  &  7  & $f_{ 2}-f_{ 1}$                                                                      &   \\
$f_{ 63}$ &  0.7847  &  0.1579  &  7  & $f_{ 18}$-2\textcolor{PineGreen}{$ f_{\rm ach1}$}                                    &   \\
$f_{ 64}$ &  1.3624  &  0.1508  &  9  &                                                                                      &   \\
$f_{ 66}$ &  2.1658  &  0.1498  &  12 &                                                                                      &   \\
$f_{ 67}$ &  0.7966  &  0.1487  &  7  & $f_{ 1}$-2\textcolor{red}{$f_{\rm rot}$}                                             &   \\
$f_{ 68}$ &  3.0239  &  0.1382  &  14 & $f_{ 1}+f_{ 16}$                                                                     &   \\
$f_{ 69}$ &  0.7986  &  0.1359  &  7  &                                                                                      &   \\
$f_{ 70}$ &  1.4300  &  0.1346  &  8  &                                                                                      &   \\
$f_{ 71}$ &  0.6503  &  0.1345  &  6  &                                                                                      &   \\
$f_{ 72}$ &  3.7452  &  0.1333  &  19 &                                                                                      &   \\
$f_{ 73}$ &  0.4637  &  0.1330  &  6  &  \textcolor{red}{$f_{\rm rot}$} - \textcolor{PineGreen}{$ f_{\rm ach2}$}             &   \\
$f_{ 76}$ &  2.2300  &  0.1276  &  10 &  $f_{ 1}$-\textcolor{PineGreen}{$ f_{\rm ach1}$}                                     &   \\
$f_{ 78}$ &  0.5274  &  0.1240  &  6  &                                                                                      &   \\
$f_{ 80}$ &  2.1607  &  0.1234  &  10 &                                                                                      &   \\
$f_{ 81}$ &  1.5056  &  0.1220  &  8  &                                                                                      &   \\
$f_{ 82}$ &  1.1604  &  0.1193  &  7  &                                                                                      &   \\
$f_{ 83}$ &  0.6317  &  0.1183  &  6  &  $2f_{ 5}-2f_{ 1}$                                                                   &   \\
$f_{ 84}$ &  1.4386  &  0.1182  &  8  &  $2f_{ 2}-2f_{ 1}$                                                                   &   \\
$f_{ 86}$ &  0.5509  &  0.1163  &  6  &                                                                                      &   \\
$f_{ 87}$ &  2.9522  &  0.1135  &  12 &                                                                                      &   \\
$f_{ 88}$ &  0.9587  &  0.1105  &  6  &  $f_{ 16}$+2\textcolor{PineGreen}{$ f_{\rm ach2}$}                                   &   \\
$f_{ 89}$ &  2.1527  &  0.1102  &  9  &  $f_{ 1}$- \textcolor{PineGreen}{$ f_{\rm ach2}$}                                    &   \\
$f_{ 91}$ &  3.6786  &  0.1078  &  16 &  \textcolor{red}{$f_{\rm rot}$}+$f_{ 2}$                                             &   \\
$f_{ 92}$ &  0.2017  &  0.1058  &  5  &                                                                                      &   \\
$f_{ 93}$ &  1.5749  &  0.1055  &  7  & $f_{ 8}$-2\textcolor{PineGreen}{$ f_{\rm ach1}$}                                     &   \\
$f_{ 94}$ &  0.9825  &  0.1048  &  6  &                                                                                      &   \\
$f_{ 95}$ &  3.0227  &  0.1022  &  11 &                                                                                      &   \\
$f_{ 98}$ &  0.9668  &  0.1007  &  6  &  $f_{ 2} -2f_{ 27}$                                                                  &   \\
$f_{ 99}$ &  1.4473  &  0.1004  &  7  &                                                                                      &   \\
$f_{ 100}$ &  0.5602  &  0.1003  &  5 &                                                                                      &   \\ 
$f_{ 102}$ &  0.9936  &  0.0992  &  6 &                                                                                      &   \\
$f_{ 103}$ &  0.4217  &  0.0985  &  5 &                                                                                      &   \\
$f_{ 105}$ &  2.9506  &  0.0982  &  11 &                                                                                     &   \\
$f_{ 109}$ &  2.2326  &  0.0953  &  8  &                                                                                     &   \\
$f_{ 111}$ &  0.1945  &  0.0935  &  5  &                                                                                     &   \\
$f_{ 112}$ &  0.5902  &  0.0932  &  5  & $3f_{ 3}-2f_{ 16}$                                                                  &   \\
$f_{ 113}$ &  1.5802  &  0.0902  &  7  &                                                                                     &   \\
$f_{ 116}$ &  0.7809  &  0.0889  &  5  &                                                                                     &   \\
$f_{ 118}$ &  2.9630  &  0.0872  &  10 &                                                                                     &    \\
$f_{ 121}$ &  2.1723  &  0.0848  &  8  & 3\textcolor{red}{$f_{\rm rot}$}+\textcolor{PineGreen}{$ f_{\rm ach1}$}              &    \\
$f_{ 122}$ &  0.2831  &  0.0841  &  5  & 3\textcolor{PineGreen}{$ f_{\rm ach1}$}+3\textcolor{PineGreen}{$ f_{\rm ach2}$}     &    \\
$f_{ 124}$ &  1.5849  &  0.0832  &  6  & $f_{ 8}$-\textcolor{PineGreen}{$ f_{\rm ach1}$}                                     &    \\
$f_{ 125}$ &  3.2129  &  0.0832  &  11 &                                                                                     &    \\
$f_{ 128}$ &  1.5737  &  0.0810  &  6  &                                                                                     &    \\
$f_{ 129}$ &  1.6460  &  0.0808  &  7  & $f_{ 2}$ +3\textcolor{PineGreen}{$ f_{\rm ach2}$} \struutdown                       &    \\
\hline
\footnotetext[1]{The Uncertainties in frequency are $\epsilon_{f}=(0.04-3) \times 10^{-4}$ d$^{-1}$.}
\footnotetext[2]{The Uncertainties in amplitude are $\epsilon_{\rm A}=(3-16) \times 10^{-3}$ mmag.}
\footnotetext[3]{Col. 'comb.' lists the combinations. For the colours in Col. 'comb.' see table~\ref{tab:terminology} and Sect.~\ref{sec:fourier-comb}. RS: rotationally split $g$ modes (Table~\ref{tab:gmode_splt} and Fig.~\ref{fig:gmode-split}.)}
\end{longtable} 
\begin{table}
 \centering
	\caption{Significant high frequencies ($f>5.$ d$^{-1}$) with the amplitudes and S/N larger than the associated mean values (A>0.0193 mmag and S/N> 7.7) among all high frequencies.  }
	\label{tab:frequencies_preigon}
	\begin{tabular}{clcccl}

\hline
$f_{\mathrm i}$ & $f$ (d$^{-1}$) & A (mmag)   & S/N & comb. & remark\struutup\struutdown\\  
\hline 
$f_{ 31}$  &  13.9651  &  0.2718  &  54 &                                                                         & \textcolor{violet}{$f_{p_{\rm max}}$}; RS; \textcolor{brown}{TDir} \struutup\\
$f_{ 39}$  &  13.0064  &  0.2219  &  53 &                                                                         & \textcolor{magenta}{$f_{\rm parent}$}; RS; TDr      \\
$f_{ 43}$  &  14.3778  &  0.2018  &  43 &                                                                         & \textcolor{magenta}{$f_{\rm parent}$}; RS; \textcolor{brown}{TDir}       \\
$f_{ 44}$  &  15.9634  &  0.1948  &  53 &                                                                         & \textcolor{magenta}{$f_{\rm parent}$}; RS; \textcolor{brown}{TDir}       \\
$f_{ 59}$  &  14.9697  &  0.1652  &  39 &                                                                         & \textcolor{magenta}{$f_{\rm parent}$}; RS; \textcolor{brown}{TDir}       \\
$f_{ 85}$  &  10.3107  &  0.1177  &  32 & 3\textcolor{magenta}{$f_{ 260}$}-2\textcolor{magenta}{$f_{ 213}$}       & TDr \\
$f_{ 104}$ &  15.9627  &  0.0985  &  30 &                                                                         & \textcolor{magenta}{$f_{\rm parent}$}; RS  \\
$f_{ 107}$ &  15.4706  &  0.0976  &  28 &                                                                         & \textcolor{magenta}{$f_{\rm parent}$}; RS; TDr \\
$f_{ 119}$ &  13.4902  &  0.0872  &  23 & 3\textcolor{PineGreen}{$ f_{\rm ach2}$}+\textcolor{magenta}{$f_{ 260}$} & \textcolor{magenta}{$f_{\rm parent}$}    \\
$f_{ 123}$ &  14.8528  &  0.0841  &  23 &                                                                         & \textcolor{magenta}{$f_{\rm parent}$}; RS; TDr \\
$f_{ 126}$ &  12.7318  &  0.0823  &  24 &                                                                         & \textcolor{magenta}{$f_{\rm parent}$}; TDr \\
$f_{ 135}$ &  16.4458  &  0.0766  &  27 &                                                                         & \textcolor{magenta}{$f_{\rm parent}$}; RS   \\
$f_{ 143}$ &  14.9677  &  0.0716  &  21 &                                                                         & \textcolor{magenta}{$f_{\rm parent}$}     \\
$f_{ 146}$ &  12.5150  &  0.0708  &  22 &                                                                         & \textcolor{magenta}{$f_{\rm parent}$}; RS; TDr  \\
$f_{ 154}$ &  14.4500  &  0.0689  &  20 &                                                                         & \textcolor{magenta}{$f_{\rm parent}$}; RS     \\
$f_{ 171}$ &  14.1831  &  0.0627  &  18 &                                                                         & \textcolor{magenta}{$f_{\rm parent}$}     \\
$f_{ 178}$ &  13.7274  &  0.0606  &  18 &                                                                         & RS; TDr \\
$f_{ 210}$ &  11.3772  &  0.0497  &  18 &                                                                         & \textcolor{magenta}{$f_{\rm parent}$}; RS  \\
$f_{ 212}$ &  15.9654  &  0.0489  &  19 &                                                                         & \textcolor{magenta}{$f_{\rm parent}$} \\
$f_{ 213}$ &  14.6942  &  0.0485  &  16 &                                                                         & \textcolor{magenta}{$f_{\rm parent}$}; RS \\
$f_{ 218}$ &  13.9660  &  0.0471  &  15 &                                                                         & \textcolor{magenta}{$f_{\rm parent}$}; RS \\
$f_{ 221}$ &  8.3516   &  0.0453  &  17 &                                                                         &  \\
$f_{ 223}$ &  13.4492  &  0.0450  &  15 &                                                                         & \textcolor{magenta}{$f_{\rm parent}$}; RS\\
$f_{ 225}$ &  10.1371  &  0.0442  &  14 &                                                                         & \\
$f_{ 226}$ &  15.4231  &  0.0441  &  16 &                                                                         & \textcolor{magenta}{$f_{\rm parent}$}; RS \\
$f_{ 228}$ &  14.8845  &  0.0438  &  15 &                                                                         & \textcolor{magenta}{$f_{\rm parent}$} \\
$f_{ 230}$ &  11.0008  &  0.0436  &  15 &                                                                         & \textcolor{magenta}{$f_{\rm parent}$}; RS \\
$f_{ 235}$ &  7.8036   &  0.0420  &  16 &                                                                         & \\
$f_{ 242}$ &  12.8819  &  0.0401  &  14 &                                                                         & \textcolor{magenta}{$f_{\rm parent}$} \\
$f_{ 246}$ &  13.3939  &  0.0391  &  13 &                                                                         & \textcolor{magenta}{$f_{\rm parent}$} \\
$f_{ 259}$ &  14.1668  &  0.0341  &  12 &                                                                         & \textcolor{magenta}{$f_{\rm parent}$}; RS \\
$f_{ 260}$ &  13.2328  &  0.0341  &  12 &                                                                         & \textcolor{magenta}{$f_{\rm parent}$}; RS\\
$f_{ 264}$ &  14.9410  &  0.0333  &  12 &                                                                         & \textcolor{magenta}{$f_{\rm parent}$} \\
$f_{ 268}$ &  14.9708  &  0.0319  &  12 & \textcolor{PineGreen}{$ f_{\rm ach2}$}+\textcolor{magenta}{$f_{ 228}$}  & \textcolor{magenta}{$f_{\rm parent}$} \\
$f_{ 269}$ &  14.3688  &  0.0318  &  11 &                                                                         & \textcolor{magenta}{$f_{\rm parent}$} \\
$f_{ 277}$ &  15.4242  &  0.0298  &  12 &                                                                         & \textcolor{magenta}{$f_{\rm parent}$}  \\
$f_{ 288}$ &  12.3408  &  0.0281  &  11 &                                                                         & RS\\
$f_{ 289}$ &  14.4482  &  0.0275  &  10 & 2\textcolor{magenta}{$f_{ 178}$}-\textcolor{magenta}{$f_{ 39}$}         & \\
$f_{ 296}$ &  14.2093  &  0.0266  &  10 & 3\textcolor{PineGreen}{$ f_{\rm ach1}$}+\textcolor{magenta}{$f_{ 171}$} & \\
$f_{ 299}$ &  12.9392  &  0.0264  &  10 &                                                                         & \\
$f_{ 302}$ &  13.9538  &  0.0256  &  9  & \textcolor{red}{$f_{\rm rot}$}+\textcolor{magenta}{$f_{260}$}           & \\
$f_{ 303}$ &  12.3687  &  0.0255  &  10 & 2\textcolor{magenta}{$f_{ 259}$}-\textcolor{magenta}{$f_{ 212}$}        & \\
$f_{ 304}$ &  14.8516  &  0.0255  &  9  &                                                                         & \\
$f_{ 317}$ &  13.9421  &  0.0231  &  9  &                                                                         & \\
$f_{ 320}$ &  20.2742  &  0.0228  &  13 &                                                                         & \\
$f_{ 322}$ &  10.4948  &  0.0223  &  8  &                                                                         &  \\
$f_{ 326}$ &  14.0506  &  0.0221  &  8  & \textcolor{violet}{$f_{p_{\rm max}}$}+\textcolor{PineGreen}{$f_{\rm ach2}$} &          \\
$f_{ 327}$ &  12.8864  &  0.0221  &  8  &                                                                         &  \\
$f_{ 328}$ &  10.8299  &  0.0218  &  8  & \textcolor{magenta}{$f_{ 230}$}-2\textcolor{PineGreen}{$ f_{\rm ach2}$} & RS \\
$f_{ 330}$ &  14.4524  &  0.0214  &  8  &                                                                         &  \\
$f_{ 331}$ &  12.0166  &  0.0213  &  8  & 3\textcolor{magenta}{$f_{ 218}$}-2\textcolor{magenta}{$f_{ 264}$}       & RS \\
$f_{ 332}$ &  13.9242  &  0.0211  &  8  &                                                                         &  \\
$f_{ 337}$ &  14.4585  &  0.0203  &  8  & \textcolor{PineGreen}{$ f_{\rm ach1}$}+\textcolor{magenta}{$f_{ 154}$}  &  \\
$f_{ 340}$ &  13.5473  &  0.0199  &  8  &                                                                         & \\
$f_{ 341}$ &  16.9162  &  0.0196  &  9  &                                                                         &   \struutdown \\
\hline
\end{tabular}
\tablefoot{ For the colours in Col. 'comb.' see table~\ref{tab:terminology} and Sect.~\ref{sec:fourier-comb}. 
{\bf RS}: {\bf R}otationally-{\bf s}plit $p$ modes (Table~\ref{tab:pmode_splt} \& Fig.~\ref{fig:pmode_splt}), 
{\bf TDr}: frequency whose {\bf T}ime {\bf D}elays show a similar long-term time delay trend (Fig.~\ref{fig:tds46fs} \& Sect.~\ref{sec:orb-new}),
{\bf TDir}: frequency whose {\bf T}ime {\bf D}elays show an incompatibility with general long-term time delay trend (Fig.~\ref{fig:tds46fs} \& Sect.~\ref{sec:orb-new}).
 For the rest of the remarks see Table~\ref{tab:terminology}.}
\end{table}
\end{appendix}
\end{document}